\documentclass[a4paper, american, english]{revtex4}

\usepackage[T1]{fontenc}
\usepackage[latin1]{inputenc}
\usepackage{xcolor}
\usepackage{soul}
\usepackage{amsthm}
\usepackage{amsmath}
\usepackage{graphicx}
\usepackage{amssymb}

\usepackage[normalem]{ulem}

\theoremstyle{plain}

\usepackage{color}
\usepackage{amsmath}

\makeatletter

\newcommand{\txt}{\textnormal}
\newcommand{\eps}{\varepsilon}

\newcommand{\emn}{\eps_{\textnormal{min}}}
\newcommand{\emx}{\eps_{\textnormal{max}}}
\newcommand{\cmx}{c_{\textnormal{{max}}}}
\newcommand{\cmn}{c_{\textnormal{{min}}}}
\newcommand{\Dmx}{\Delta_{\textnormal{max}}}
\newcommand{\Dmn}{\Delta_{\textnormal{min}}}

\newcommand{\Post}{\textnormal{Post}}
\newcommand{\Pre}{\textnormal{Pre}}

\newcommand{\tsf}{\txt{s}}
\newcommand{\trf}{\txt{r}}
\newcommand{\free}{\txt{free}}

\bibpunct{(}{)}{,}{a}{,}{,}

\usepackage{babel}

\begin{document}

\title{How Chaotic is the Balanced State?}

\author{Sven Jahnke$^{1,2}$, Raoul-Martin Memmesheimer$^{1-3}$ and Marc
Timme$^{1,2}$}

\selectlanguage{english}%

\affiliation{$^{1}$Network Dynamics Group, Max-Planck-Institute for Dynamics
\& Self-Organization (MPIDS), }

\affiliation{$^{2}$Bernstein Center for Computational Neuroscience (BCCN), 37073
G\"ottingen, Germany,}

\affiliation{$^{3}$Center for Brain Science, Faculty of Arts and Sciences Harvard
University, Cambridge, MA02138, USA}
\selectlanguage{american}%

\begin{abstract}
Large sparse circuits of spiking neurons exhibit a balanced state
of highly irregular activity under a wide range of conditions. It
occurs likewise in sparsely connected random networks that receive
excitatory external inputs and recurrent inhibition as well as in
networks with mixed recurrent inhibition and excitation. Here we analytically
investigate this irregular dynamics in finite networks keeping track
of all individual spike times and the identities of individual neurons.
For delayed, purely inhibitory interactions we show that the irregular
dynamics is not chaotic but in fact stable. Moreover, we demonstrate
that after long transients the dynamics converges towards periodic
orbits and that every generic periodic orbit of these dynamical systems
is stable. We investigate the collective irregular dynamics upon increasing
the time scale of synaptic responses and upon iteratively replacing
inhibitory by excitatory interactions. Whereas for small and moderate
time scales as well as for few excitatory interactions, the dynamics
stays stable, there is a smooth transition to chaos if the synaptic
response becomes sufficiently slow (even in purely inhibitory networks)
or the number of excitatory interactions becomes too large. These
results indicate that chaotic and stable dynamics are equally capable
of generating the irregular neuronal activity. More generally, chaos
apparently is not essential for generating high irregularity of balanced
activity, and we suggest that a mechanism different from chaos and
stochasticity significantly contributes to irregular activity in cortical
circuits.
\end{abstract}

\keywords{balanced state, irregular activity, local cortical circuits, synchronization,
attractor neural networks, stability}

\maketitle

\section{Introduction}

Most neurons in the brain communicate by emitting and receiving electrical
pulses, called action potentials or spikes, via chemically operating
synaptic connections. Local cortical circuits often exhibit spiking
dynamics that is highly irregular and appears as if it were random.
Such irregular activity at low neuronal firing rate is thus considered
a basic ``ground state''. It is characterized by individual neurons
that display largely fluctuating membrane potentials and highly variable
inter-spike-intervals (ISIs) as well as by low correlations between
the neurons \citep{Vreeswijk1996,Vreeswijk1998,Brunel2000,Vogels2005,Kumar2007}.
Originally, this dynamical state seemed to be in contradiction to cortical
anatomy, where each neuron receives a huge number of synapses, typically
$10^{3}-10^{4}$ \citep{Braitenberg1998}: One might expect that a
large number of uncorrelated, or weakly correlated synaptic inputs
to one neuron, given the central limit theorem, sums up to a regular
total input signal with only small relative fluctuations, therefore
excluding the emergence of irregular dynamics. So the finding of highly
irregular activity might be surprising.

This issue was resolved by the idea of a ``balanced state'' \citep{Vreeswijk1996},
in which excitatory (positive) and inhibitory (negative) input balances
such that the average membrane potential is sub-threshold and strong
fluctuations once in a while are sufficiently depolarizing to initiate
a spike. The original work ``Chaos in neuronal networks with balanced
excitatory and inhibitory activity'' \citep{Vreeswijk1996} was an
analysis of a self-consistent, highly irregular ``balanced'' activity
for sparse random networks of binary model neurons. It was shown that
the balanced state naturally and robustly occurs in large networks
if the inhibitory and excitatory coupling strengths and their respective
numbers of synapses appropriately scale with each other. Moreover,
flipping the state of only one of the binary neurons in a large network
(i.e. applying the smallest possible non-zero perturbation in such
a system) leads to a supra-exponential divergence between the perturbed
and the unperturbed realizations of the network dynamics, exemplifying
the extremely chaotic nature of the balanced activity \citep{Vreeswijk1996,Vreeswijk1998}. 

\begin{figure*}
\begin{centering}
\includegraphics[clip,width=140mm]{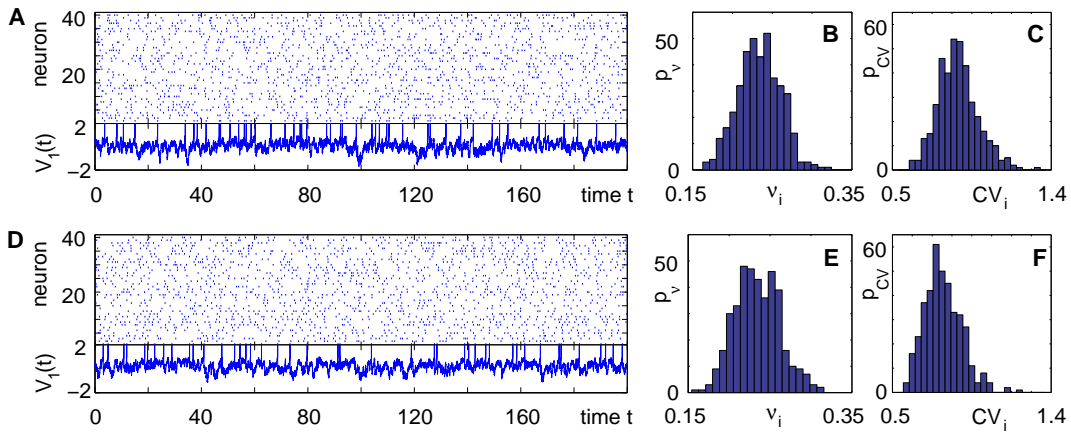}
\par\end{centering}

\caption{Highly irregular spiking activity equally emerges from chaotic and
from stable circuit dynamics. Irregular dynamics in purely inhibitorily
(A-C) and inhibitorily and excitatorily (D-F) coupled random networks
of identical leaky integrate-and-fire neurons ($N=400$, $\gamma_{i}\equiv1$,
$V_{\Theta,i}\equiv1$, $\tau_{ij}\equiv0.1T_{i}^{\free}$, $\left|\Pre(i)\right|\equiv80$).
(A,D) Spiking dynamics (A)$\sum_{j}\eps_{i,j}\equiv-16$, $I_{i}\equiv4$,
$N_{E}=0$; (D)$\sum_{j}\eps_{i,j}\equiv-11$, $I_{i}\tilde{\equiv}2.7$,
$N_{E}=1000$, where the $N_{E}$ excitatory couplings are distributed
such that each neuron has the same number of excitatory inputs. The
upper panel displays the spiking times (blue lines) of the first $40$
neurons. The lower panel displays the membrane potential trajectory
of neuron $i=1$ (spikes of height $\Delta V=1$ added at firing times).
(B,E) Histogram of mean firing rates $\nu_{i}$. (C,F) Histogram of
the coefficients of variation $\textsf{CV}_{i}:=\sigma_{i}/\mu_{i}$;
$\mu_{i}=\left\langle t_{i,k+1}^{\tsf}-t_{i,k}^{\tsf}\right\rangle _{k}$;
$\sigma_{i}^{2}:=\left\langle \left(t_{i,k+1}^{\tsf}-t_{i,k}^{\tsf}-\mu_{i}\right)^{2}\right\rangle _{k}$
averaged over time. \label{fig:balanced}}

\end{figure*}

Later analysis \citep{Amit1997,Brunel2000} of networks of integrate-and-fire
neurons demonstrated that the mean field description of the balanced
activity for these continuous-state neuron networks is very similar
to that of the original binary neuron networks. These and related
results \citep{Amit1997,Brunel2000,Timme2002b,Timme2008} indicate
that statistically the same balanced activity persists both in networks
with external excitatory inputs and recurrent inhibition only as well
as in networks with equal total amounts of recurrent inhibition and
recurrent excitation. These findings about the robustness of the balanced
state, the original work \citep{Vreeswijk1996,Vreeswijk1998} together
with common intuition may suggest that highly irregular activity originates
from chaotic network dynamics. This hypothesis, however, has not been
systematically investigated so far. Recent research even points towards
the contrary: it shows that in globally coupled networks without delay
the dynamics tends to converge to stable periodic orbits if inhibition
dominates \citep{Jin2002}. Numerical investigations of weakly diluted
inhibitorily coupled networks without delay even show that although
the dynamics may be irregular, its Lyapunov exponent is negative \citep{Zillmer2006}.
These numerical simulations also demonstrate by example that the dynamics
converges to a periodic orbit after long quasi-stationary transients.
Interestingly, a related article \citep{Zillmer2009}
also gives numerical evidence that there can be chaos and long transients
even in networks with only inhibitory connections.

In this article we show analytically in the limit of fast synaptic
response, that in inhibitory networks with inhomogeneous delay distribution
and arbitrary, strongly connected topology (A network is {\it{strongly connected}} if there is a directed path of connections between any ordered pair of neurons.) any generic trajectory is asymptotically stable. After a (typically long) stable transient characterized by irregular activity the dynamics
converges to a periodic orbit that is also stable, in agreement with
the results presented in \citep{Memmesheimer2006a,Timme2008}. In
particular the transients are not chaotic in contrast to the ones
occurring in purely excitatorily coupled networks \citep{Zumdieck2004}.
We show that this collective dynamics is robust upon increasing the
synaptic response time from zero and upon replacing some inhibitory
by excitatory interactions. Nevertheless, if the synaptic response
becomes too slow or the number of excitatory interaction too large,
the collective dynamics becomes chaotic via a transition where the
Lyapunov exponent changes smoothly and the spiking activity stays
highly irregular. Thus the irregularity equally prevails in networks
with stable as well as in networks with chaotic dynamics, leaving
no evidence that chaos generates the irregularity.

Some analytically accessible aspects of the stable irregular dynamics
have been briefly reported before \citep{Jahnke2008}. Parts of this
work have been presented in \citep{Memmesheimer2007} and at a Bernstein
Symposium \citep{Jahnke2008BCCN}.

\section{Network model\label{sec:Network-model}}

We consider networks of $N$ neurons with directed couplings that
interact by sending and receiving spikes. If such a directed connection
exists from neuron $j$ to neuron $i$, we call $i$ postsynaptic
to neuron $j$. We denote the set of all postsynaptic neurons of neuron
$j$ by $\Post{(j)}$. Neuron $j$ is then presynaptic to neuron $i$,
the set of all presynaptic neurons is denoted by $\Pre(i)$. The membrane
potential $V_{i}(t)$ of some neuron $i$ evolves according to \begin{equation}
\frac{d}{dt}V_{i}=f_{i}(V_{i})+\sum_{j=1}^{N}\sum_{k\in\mathbb{Z}}\eps_{ij}\delta\left(t-t_{jk}^{\tsf}-\tau_{ij}\right),\label{eq:Vdot}\end{equation}
where a smooth function $f_{i}$ specifies the internal dynamics,
$\eps_{ij}$ are the coupling strengths from presynaptic neurons $j$
to $i$, $\delta(.)$ is the Dirac delta-distribution, $\tau_{ij}>0$
are the delay times of the connection and $t_{jk}^{\tsf}$ denotes
the time when neuron $j$ sends the $k$th spikes. If not stated otherwise,
in the following we consider inhibitory networks, i.e. $\eps_{ij}\le0$.
Spikes sent to postsynaptic neurons with different delay times from
neuron $j$ are considered as separate spikes, so if the delays are
all different, e.g.~if they are chosen randomly, neuron $j$ sends
$\left|\Post(j)\right|$ spikes at time $t_{jk}^{\tsf}$. When neuron
$j$ reaches the threshold $V_{\Theta,j}$ of the potential, i.e.~$V_{j}(t^{-})=V_{\Theta,j}$,
it generates spikes at $t=:t_{jk}^{\tsf}$ for some $k$ and is reset,
$V_{j}(t_{jk}^{\tsf})=0$. The neuronal dynamics is therefore smooth
except at times when events, namely sendings or receivings of spikes
happen. Simultaneous sendings of spikes by one neuron are treated
as one event as well as simultaneous receptions of spikes sent by
the same neuron. We require that the $f_{j}$ satisfy $f_{j}(V_{j})>0$
and $f_{j}{}^{\prime}(V_{j})<0$ for all $j$ and $V_{j}\leq V_{\Theta,j}$,
such that in isolation each neuron $j$ exhibits oscillatory dynamics
with a period $T_{j}^{\free}$.

\begin{figure*}
\begin{centering}
\includegraphics[clip,width=140mm]{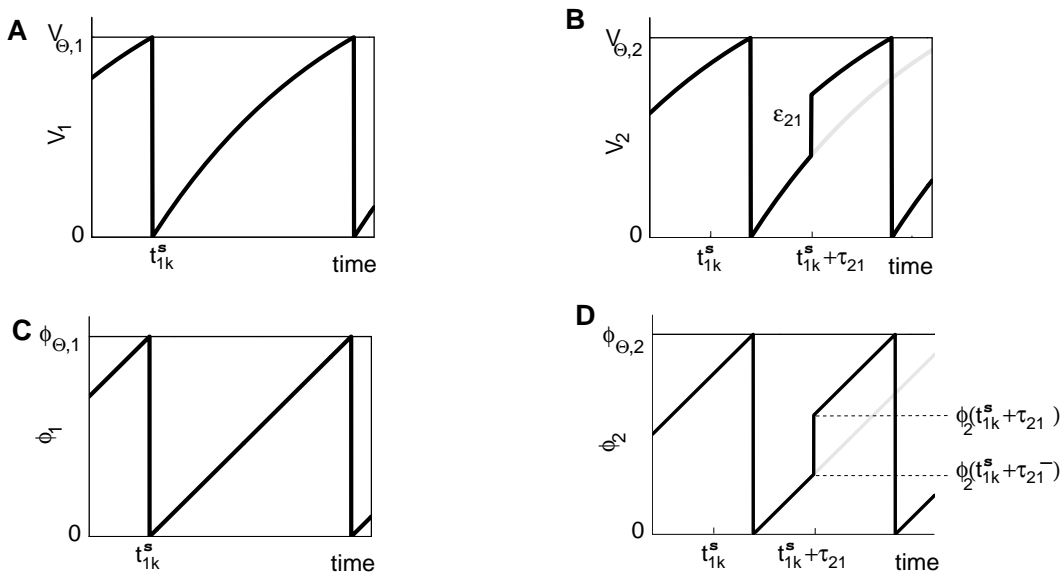}
\par\end{centering}

\caption{Relation between membrane potential and phase dynamics. (A) Membrane
potential of neuron $1$. At $t=t_{1k}^{\tsf}$ the membrane potential
$V_{1}(t)$ crosses the threshold potential $V_{\Theta,1}$ which
leads to a reset of the potential to $V_{1}(t_{1k}^{\tsf})=0$, and
spikes are emitted. (B) Membrane potential of neuron $2$ which is
postsynaptic to neuron $1$, i.e.~$2\in\Post(1)$. The spike is received
at $t_{1k}^{\tsf}+\tau_{21}$ and induces a jump of size $\eps_{21}$
in the potential. (C) Phase dynamics of neuron $1$. The phase increases
linearly until the phase threshold $\phi_{\Theta,1}$ is reached,
then it is set to zero. (D) Phase dynamics of neuron $2$. When the
spike is received at $t=t_{1k}^{\tsf}+\tau_{21}$ it induces a phase
jump: $\phi_{2}(t_{1k}+\tau_{21})=H_{\eps_{21}}^{(2)}(\phi_{2}(t_{1k}+\tau_{21}))=U_{2}^{-1}(U_{2}(\phi_{2}((t_{1k}+\tau_{21})^{-}))+\eps_{21})$.\label{fig:potential_phases}}

\end{figure*}

The network dynamics can equivalently be described by a phase-like
variable $\phi_{j}(t)\in\left(-\infty,\phi_{\Theta,j}\right]$ satisfying
\begin{equation}
d\phi_{j}/dt=1\label{eq:phidot}\end{equation}
at all non-event times \citep{Mirollo1990}. When the phase threshold
is reached, $\phi_{j}({t_{jk}^{\tsf}}^{-})=\phi_{\Theta,j}$, the
phase is reset, $\phi_{j}(t_{jk}^{\tsf}):=0$ and a spike is generated.
This spike travels to the postsynaptic neurons, arrives after a delay
time $\tau_{ij}$ at neuron $i$ and induces a phase change according
to 

\begin{equation}
\phi_{i}(t_{jk}^{\tsf}+\tau_{ij})=H_{\eps_{ij}}^{(i)}\left(\phi_{i}\left(\left(t_{jk}^{\tsf}+\tau_{ij}\right)^{-}\right)\right).\label{eq:phiupdate}\end{equation}
with the transfer function \begin{equation}
H_{\eps}^{(i)}(\phi):=U_{i}^{-1}\left[U_{i}(\phi)+\eps\right],\label{eq:H}\end{equation}
where each $U_{i}(t)$ is the free (all $\eps_{ij}=0$) solution of
(\ref{eq:Vdot}) through the initial condition $U_{i}(0)=0$, yielding
$U_{i}'>0$ and $U_{i}''<0$ and $\phi_{\Theta,j}=U_{j}^{-1}(V_{\Theta,j})$,
cf. \citep{Memmesheimer2006a}. Fig.~\ref{fig:potential_phases}
illustrates the relation between phase dynamics and membrane potential. 

The analysis below is valid for general $U_{i}(\phi)$; in the numerical
simulations we employ leaky integrate-and-fire neurons, $f_{i}(V):=I_{i}-\gamma_{i}V$
with time scale $\gamma_{i}^{-1}>0$ and equilibrium potential $\gamma_{i}^{-1}I_{i}>V_{\Theta,i}$,
the membrane potential has the functional dependence $U_{i}(\phi)={\gamma_{i}^{-1}I}_{i}(1-\exp(-\gamma_{i}\phi))$
on $\phi$ and the oscillation period of a free neuron is given by
$T_{i}^{\free}:=\gamma_{i}^{-1}\ln(I_{i}/(I_{i}-\gamma_{i}V_{\Theta,i}))$.
We consider arbitrary generic spike sequences in which all neurons
are active, i.e. there is a finite constant $T>0$, such that in every
time interval $[t,t+T)$, $t\in\mathbb{R}$, every neuron fires at
least once. Further, we assume that the dynamics is sufficiently irregular
such that two events occur at the same time with zero probability.

Due to the delay, the state space is formally infinite dimensional. However, it becomes finite dimensional after some finite time $t^{\prime}$ (cf.~\citep{Ashwin2005}).
At a given time $t>t^{\prime}$ the network dynamics is completely
determined by the phases $\phi_{i}(t)$ and by the spikes which have
been sent but not yet received by the postsynaptic neurons at $t$.
Their number is bounded by some constant $ND'$. Due to the inhibitory
character of the network couplings, each neuron $i$ needs at least
the free oscillation period $\phi_{\Theta,i}=T_{i}^{\free}$ to generate
a spike after the last reset. Consequently, at most\begin{equation}
D^{\prime}=N\left\lceil \max_{i,j\in\{1,\ldots,N\}}\left(\frac{\tau_{ij}}{\phi_{\Theta,i}}\right)\right\rceil \label{eqn:SpikesOnTheWayPerNeuron}\end{equation}
spikes per neuron are in transit and the state space stays finite,
with dimensionality smaller than or equal to $N\cdot(1+D^{\prime})$
(cf.~also \citep{Ashwin2005}). Here$\left\lceil x\right\rceil $
denotes the ceiling function, the smallest integer larger or equal
to $x$.

We now introduce variables to describe spikes, which are already sent
at time $t$ by neuron $j$ to the postsynaptic neuron $i$ and not
yet received. A single spike in transit is characterized by the state
variable $\sigma_{ijk}(t)\in\left[0,\tau_{ij}\right]$. The index
$k=1,2,3\ldots\leq D^{\prime}/N$ numbers the different spikes traveling
from neuron $j$ to $i$ at time $t$ in the order of arrival at the
postsynaptic neuron $i$, starting with $k=1$ for the next spike
to arrive. When spikes are emitted at time $t_{jn}^{\tsf}$ for some
$n$, $\sigma_{ijk}(t)$ is set to $\sigma_{ijk}(t_{jn}^{\tsf})=0$.
The spike index $k$ equals the number of spikes already in transit
plus one. It thus depends on the actual network state at time $t_{jn}^{\tsf}$.
Between two events $\sigma_{ijk}(t)$ increases linearly with slope
one, when $\sigma_{ijk}(t_{jn}^{\tsf}+\tau_{ij})=\tau_{ij}$ the spike
is received by the postsynaptic neuron $i$, where it induces a phase
jump according to Eq.~\eqref{eq:phiupdate}. After spike reception
we cancel the spike arrived (which has index $k=1$) and renumber
the indices $k>1$ as $k\rightarrow k-1$ such that $\sigma_{ij1}(t_{jn}^{\tsf}+\tau_{ij})$
specifies the spike sent by neuron $j$ which arrives next at the
postsynaptic neuron $i$ (cf. Fig.~\ref{fig:defDeltas}(A,B) for
illustration). 

\section{Results}

\subsection{Lyapunov stability of arbitrary generic spike sequences}

In this section we study the stability properties of the spike sequences.
We compare the microscopic dynamics of two sequences, that slightly
differ in the timing of the spikes, but have the same ordering. We
show that the distance between these trajectories is bounded by the
initial distance. Assuming that one sequence is generated by a perturbation
of the other, this implies Lyapunov stability for the considered spike
pattern. Distances and perturbation sizes are measured using the maximum
norm.

\begin{figure*}
\begin{centering}
\includegraphics[clip,width=140mm]{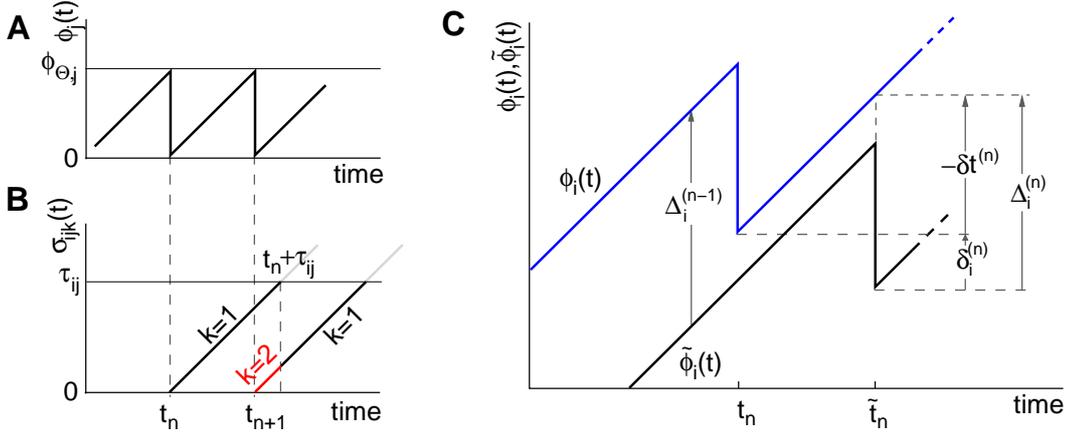}
\par\end{centering}

\caption{(A) Phase dynamics of neuron $j$ and definition of state variables.
At $t=t_{n}$ and $t=t_{n+1}$ the phase $\phi_{j}(t)$ reaches the
threshold and is reset to $0$. (B) The spikes emitted travel to the
postsynaptic neurons $i\in\Post(j)$. They are described by $\sigma_{ijk}(t)$.
In this example we show two spikes traveling from neuron $j$ to one
specific postsynaptic neuron $i,$ described by $\sigma_{ij1}(t)$
(black) and $\sigma_{ij2}(t)$ (red). At $t=t_{n}$, $\sigma_{ij1}(t_{n})$
is set to $0$; here $k=1$ because there is no spike in transit at
$t=t{}_{n}^{-}$. When neuron $j$ spikes again at $t=t_{n+1}$, $\sigma_{ij2}(t_{n+1})$
is set to $0$. At $t=t_{n}+\tau_{ij}$ the spike emitted at $t=t_{n}$
arrives at the postsynaptic neuron $i$ and induces a phase jump in
$\phi_{i}(t)$ (not shown, cf. Fig.~\ref{fig:potential_phases}).
After spike reception, we renumber $k\rightarrow k-1$ such that $\sigma_{ij2}\rightarrow\sigma_{ij1}$.
(C) Definition of the phase shifts. The phase curve $\phi_{i}(t)$
of neuron $i$ (blue) before and after the reception of a spike at
$t=t_{n}$ is shown together with $\tilde{\phi}_{i}(t)$ (black).
$\delta_{i}^{(n)}=\phi_{i}\left(t_{n}\right)-\tilde{\phi}_{i}\left(\tilde{t}_{n}\right)$
is the difference of neuron $i$'s phases in the unperturbed and the
perturbed dynamics taken at corresponding event times $t_{n}$ and
$\tilde{t}_{n}$. $\delta t^{(n)}=t_{n}-\tilde{t}_{n}$ denotes the
difference of event times $t_{n}$ and $\tilde{t}_{n}$, i.e.~the
temporal offset between both sequences. Finally, $\Delta_{i}^{(n)}=\delta_{i}^{(n)}-\delta t^{(n)}$
is some phase shift of neuron $i$ with the temporal offset taken
into account. \label{fig:defDeltas}}

\end{figure*}
Since the distance between trajectories only changes at event times,
we can choose an event-based perspective. The time of the $n$th event
in the entire network is denoted by $t_{n}$ in the first sequence,
and by $\tilde{t}_{n}$ in the second one. Analogously, we denote
the phases of a neuron $i$ at a given time $t$ by $\phi_{i}(t)$
and $\tilde{\phi}_{i}(t)$ and the spikes in transit at time $t$
by $\sigma_{ijk}(t)$ and $\tilde{\sigma}_{ijk}(t)$ in the different
sequences. Let\begin{equation}
\Delta_{i}^{(n)}:=\left(\phi_{i}\left(t_{n}\right)-\tilde{\phi}_{i}\left(\tilde{t}_{n}\right)\right)-\left(t_{n}-\tilde{t}_{n}\right)=\delta_{i}^{(n)}-\delta t^{(n)}\label{eqn:DefinitionDeltaPhi}\end{equation}
denote the difference between the phase difference, $\delta_{i}^{(n)}:=\phi_{i}\left(t_{n}\right)-\tilde{\phi}_{i}\left(\tilde{t}_{n}\right)$,
and the temporal offset, $\delta t^{(n)}:=t_{n}-\tilde{t}_{n}$, after
the $n$th and before the $(n+1)$th event (cf.~Fig.~\ref{fig:defDeltas}(C)).
Similarly \begin{equation}
{\Delta\sigma}_{ijk}^{(n)}:=\left(\sigma_{ijk}\left(t_{n}\right)-\tilde{\sigma}_{ijk}\left(\tilde{t}_{n}\right)\right)-\left(t_{n}-\tilde{t}_{n}\right)={\delta\sigma}_{ijk}^{(n)}-\delta t^{(n)}\label{eqn:DefinitionDeltaSigma}\end{equation}
labels the shift of the $k$th spike sent by neuron $j$ and not yet
arrived at neuron $i$ after the $n$th and before the $(n+1)$th
event. Between two consecutive events, both $\phi_{i}(t),\tilde{\phi}_{i}(t)$
and $\sigma_{ijk}(t),\tilde{\sigma}_{ijk}(t)$ increase linearly and
only at event times the phases and spike variables are updated nonlinearly
as described above. Therefore, to study the stability of the system
it is sufficient to consider the phase shifts after events.

In the following we investigate the evolution of shifts at the discrete
event times. There are two different kinds of events: (i) sending
and (ii) receiving of spikes. In the first case, the shifts $\Delta_{i}^{(n)}$
and $\Delta\sigma_{ijk}^{(n)}$ stay unchanged, but new spikes with
new spike variables are generated. These variables inherit the perturbation
of the sending neuron. In the second case, the phase shift of the
neuron which receives the spike changes and the spikes in transit
are reordered. The resulting phase shift of the neuron receiving a
spike turns out to be a weighted sum of previous shifts. This can
be shown by studying both cases in detail.
\begin{enumerate}
\item \textbf{Transfer of perturbations without change of size. }If as $(n+1)$th
event the phase of some neuron $j^{\ast}$ reaches its threshold and
a spike is emitted, the shifts of all neurons' phases stay unchanged,\begin{equation}
\Delta_{j}^{(n+1)}=\phi_{i}\left(t_{n}\right)+\left(t_{n+1}-t_{n}\right)-\left(\tilde{\phi}_{i}\left(\tilde{t}_{n}\right)+\left(\tilde{t}_{n+1}-\tilde{t}_{n}\right)\right)-\delta t^{(n+1)}=\Delta_{j}^{(n)}.\label{eqn:UpdateDeltaPhiOnSending}\end{equation}
Similarly, the shifts of the spikes in transit stay unchanged\begin{equation}
\Delta\sigma_{ijk}^{(n+1)}=\sigma_{ijk}\left(t_{n}\right)+\left(t_{n+1}-t_{n}\right)-\left(\tilde{\sigma}_{ijk}\left(\tilde{t}_{n}\right)+\left(\tilde{t}_{n+1}-\tilde{t}_{n}\right)\right)-\delta t^{(n+1)}={\Delta\sigma}_{ijk}^{(n)}.\label{eqn:UpdateDeltaSigmaOnSending}\end{equation}
Additionally new spikes are generated $\sigma_{ij^{\ast}k^{\ast}}(t_{n+1})=0$
and $\tilde{\sigma}_{ij^{\ast}k^{\ast}}(\tilde{t}{}_{n+1})=0$ where
$k^{\ast}=k^{\ast}(i,j^{\ast},n+1)$ is the appropriate spike number,
cf.~Fig.~\ref{fig:defDeltas}. The shifts of the new spike variables
depend on the phase shift of the sending neuron $j^{\ast}$according
to\begin{equation}
{\Delta\sigma}_{ij^{\ast}k^{\ast}}^{(n+1)}=-\delta t^{(n+1)}=-\left[\left(t_{n}+\phi_{\Theta,j^{\ast}}-\phi_{j^{\ast}}\left(t_{n}\right)\right)-\left(\tilde{t}{}_{n}+\phi_{\Theta,j^{\ast}}-\tilde{\phi}_{j^{\ast}}\left(\tilde{t}{}_{n}\right)\right)\right]=\Delta_{j^{\ast}}^{(n)}.\label{eq:ShiftOfnewSpike}\end{equation}

\item \textbf{Averaging of prior perturbations. }If as $(n+1)$th event
some spike arrives, say $\sigma_{i^{\ast}j^{\ast}1}\left(t_{n+1}\right)=\tau_{i^{\ast}j^{\ast}}$,
it induces a phase jump in the postsynaptic neuron $i^{\ast}$. According
to Eq.~\eqref{eq:phiupdate}, the phase shift $\Delta_{i^{\ast}}^{(n+1)}$
can be computed as\begin{equation}
\Delta_{i^{\ast}}^{(n+1)}=H_{\eps_{i^{\ast}j^{\ast}}}^{(i^{\ast})}\left(\phi_{i^{\ast}}\left(t_{n+1}^{-}\right)\right)-H_{\eps_{i^{\ast}j^{\ast}}}^{(i^{\ast})}\left(\tilde{\phi}{}_{i^{\ast}}\left({\tilde{t}_{n+1}}^{-}\right)\right)-\delta t^{(n+1)},\label{eqn:DeltaNplusOne}\end{equation}
where $\phi_{i^{\ast}}\left(t_{n+1}^{-}\right)=\phi_{i^{\ast}}(t_{n})+\left(t_{n+1}-t_{n}\right)$
and $\tilde{\phi}_{i^{\ast}}\left(\tilde{t}_{n+1}^{-}\right)=\tilde{\phi}_{i^{\ast}}(\widetilde{t}_{n})+\left(\widetilde{t}_{n+1}-\widetilde{t}_{n}\right)$
are the phases {}``just before'' spike reception. Using the definitions
\eqref{eqn:DefinitionDeltaPhi} we find the identity \begin{equation}
\phi_{i^{\ast}}\left({t_{n+1}}^{-}\right)=\tilde{\phi}_{i^{\ast}}\left({\tilde{t}_{n+1}}^{-}\right)+\Delta_{i^{\ast}}^{(n)}+\delta t^{(n+1)}.\end{equation}
 Applying the mean value theorem in Eq.~\eqref{eqn:DeltaNplusOne}
and the relation\begin{align}
\delta t^{(n+1)}= & t_{n+1}-\tilde{t}_{n+1}\nonumber \\
= & t_{n}+\tau_{i^{*}j^{*}}-\sigma_{i^{\ast}j^{\ast}1}(t_{n})-\tilde{t}_{n}-\tau_{i^{*}j^{*}}+\tilde{\sigma}_{i^{\ast}j^{\ast}1}(\tilde{t}_{n})\nonumber \\
= & -\Delta\sigma_{i^{\ast}j^{\ast}1}^{(n)}\label{eq:-1}\end{align}
 yields \begin{equation}
\Delta_{i^{\ast}}^{(n+1)}=c_{i^{\ast}j^{\ast}}^{(n+1)}\cdot\Delta_{i^{\ast}}^{(n)}+\left(1-c_{i^{\ast}j^{\ast}}^{(n+1)}\right)\cdot\Delta\sigma_{i^{\ast}j^{\ast}1}^{(n)},\label{eq:Averaging}\end{equation}
where $c_{i^{\ast}}^{(n+1)}$ is given by the derivative \begin{equation}
c_{i^{\ast}j^{\ast}}^{(n+1)}=\left.\frac{\partial H_{\eps_{i^{\ast}j^{\ast}}}^{(i^{\ast})}\left(\phi\right)}{\partial\phi}\right|_{\phi\in\left[\phi_{i^{\ast}}\left(t_{n+1}^{-}\right),\tilde{\phi}_{i^{\ast}}\left(\tilde{t}_{n+1}^{-}\right)\right]}\end{equation}
for $\phi_{i^{\ast}}\left(t_{n+1}^{-}\right)\leq\tilde{\phi}_{i^{\ast}}\left(\tilde{t}_{n+1}^{-}\right)$;
for $\phi_{i^{\ast}}\left(t_{n+1}^{-}\right)>\tilde{\phi}_{i^{\ast}}\left(\tilde{t}_{n+1}^{-}\right)$,
$\phi$ takes values $\phi\in\left[\tilde{\phi}_{i^{\ast}}\left(\tilde{t}_{n+1}^{-}\right),\phi_{i^{\ast}}\left(t_{n+1}^{-}\right)\right]$.
If neuron $i^{\ast}$ is not connected to neuron $j^{\ast}$, $\eps_{i^{\ast}j^{\ast}}=0$,
the function $H_{\eps_{i^{\ast}j^{\ast}}}^{(i^{\ast})}\left(\phi\right)=H_{0}^{(i^{\ast})}(\phi)=\phi$
becomes the identity, such that the phase shift stays unchanged, $\Delta_{i^{\ast}}^{(n+1)}=\Delta_{i^{\ast}}^{(n)}$,
indeed $c_{i^{\ast}j^{\ast}}^{(n+1)}=dH_{0}^{(i^{\ast})}(\phi)/d\phi=1$
independent of $\phi$.\\
For $\eps_{i^{\ast}j^{\ast}}<0$, $c_{i^{\ast}}^{(n+1)}$ is bounded
by\begin{equation}
\cmn:=\inf_{\phi,k}\left\{ \frac{\partial\left(H_{\emn}^{(k)}\left(\phi\right)\right)}{\partial\phi}\right\} \leq c_{i^{\ast}}^{(n+1)}\leq\sup_{\phi,k}\left\{ \frac{\partial\left(H_{\emx}^{(k)}\left(\phi\right)\right)}{\partial\phi}\right\} =:\cmx,\label{eq:DefCmnCmx}\end{equation}
where $\emx:=\max\limits _{i,j:\eps_{ij}\neq0}\left\{ \eps_{ij}\right\} $
and $\emn:=\min\limits _{i,j:\eps_{ij}\neq0}\left\{ \eps_{ij}\right\} $.
We have used that $\partial\left(H_{\eps}^{(k)}\left(\phi\right)\right)/\partial\phi$
is monotonic increasing with $\eps$\begin{eqnarray}
\frac{\partial}{\partial\eps}\left\{ \frac{\partial H_{\eps}^{(k)}(\phi)}{\partial\phi}\right\}  & = & U_{k}^{\prime}(\phi)\frac{\partial}{\partial\eps}\left(\left[U_{k}^{\prime}(H_{\eps}^{(k)}(\phi))\right]^{-1}\right)\\
 & = & -\frac{U_{k}^{\prime}(\phi)}{\left(U_{k}^{\prime}(H_{\eps}^{(k)}(\phi))\right)^{2}}\cdot U_{k}^{\prime\prime}(H_{\eps}^{(k)}(\phi))\cdot\frac{1}{U_{k}^{\prime}(H_{\eps}^{(k)}(\phi))}>0.\label{eqn:MonotonicityInEpsilon}\end{eqnarray}
\begin{figure}
\begin{centering}
\includegraphics[clip,width=85mm]{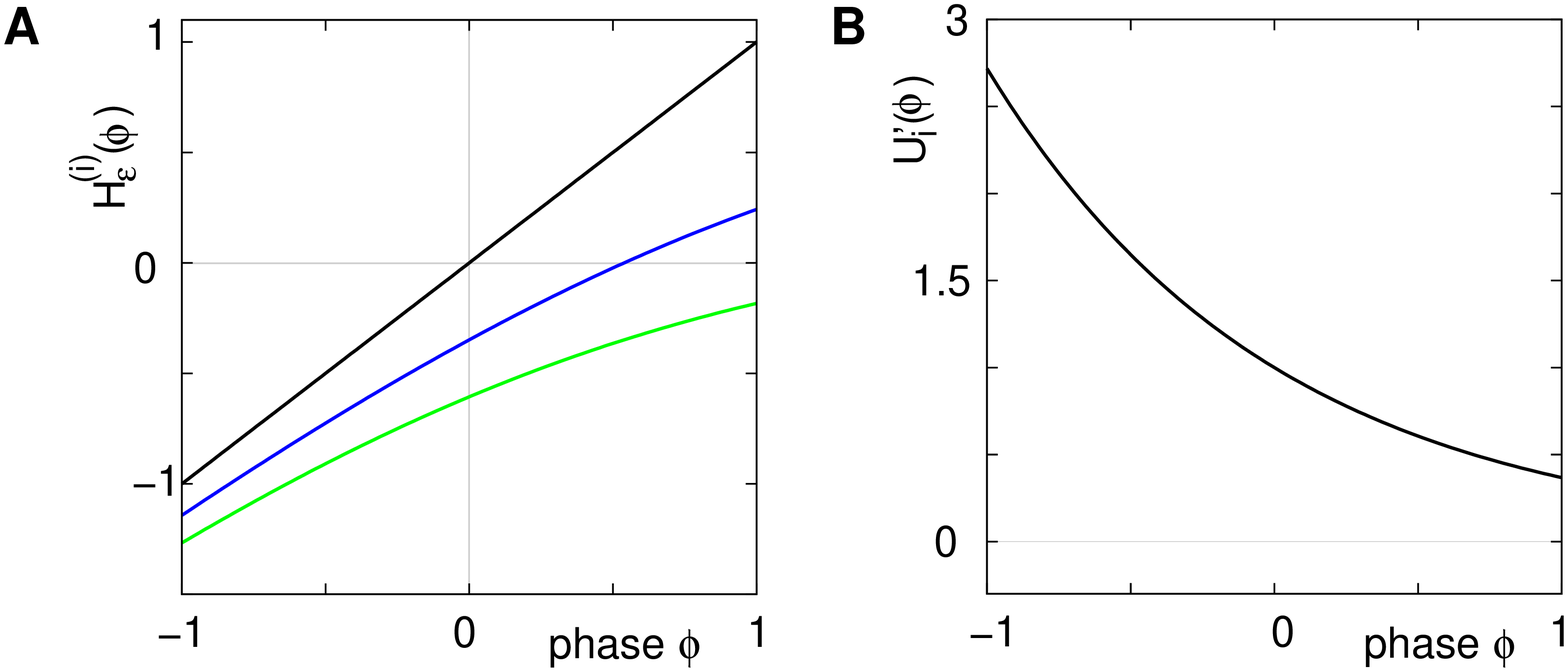}
\par\end{centering}

\caption{(A) The transfer function $H_{\eps}^{(i)}(\phi)$ for a leaky
integrate-and-fire neuron. For $\eps=0$ (black), $H_{0}^{(i)}(\phi)=\phi$
is the identity (black), for $\eps<0$ (blue: $\eps=-0.5$, green:
$\eps=-1$) the phase $\phi$ after receiving an input is smaller
than before. For all inhibitory inputs we find $H_{\eps}^{(i)}(\phi)<\phi.$
(B) The derivative of $U_{i}(\phi)$ is monotonic decreasing, therefore
$\partial H_{\eps}^{(i)}(\phi)/\partial\phi\leq1$ for $\eps\leq0$
(cf.~Eq.~\eqref{eq:boundingofHprime}).\label{fig:visualH}}

\end{figure}
 \\
The shifts of traveling spikes stay unchanged on spike reception,
cf.~Eq.~\eqref{eqn:UpdateDeltaSigmaOnSending}, \begin{equation}
\Delta\sigma_{ijk}^{(n+1)}=\Delta\sigma_{ijk}^{(n)}\end{equation}
for all spikes with $i\neq i^{\ast}\vee j\neq j^{\ast}$. For ${i=i}^{\ast},j=j^{\ast}$
the spike variables are renumbered and\begin{equation}
\Delta\sigma_{i^{\ast}j^{\ast}k-1}^{(n+1)}=\Delta\sigma_{i^{\ast}j^{\ast}k}^{(n)}\end{equation}
holds, except for $k=1$, because $\sigma_{i^{\ast}j^{\ast}1}(t)$
is the variable describing the spike received and therefore canceled.
\end{enumerate}
We will now ascertain that the coefficients $c_{i^{\ast}j^{\ast}}^{(n+1)}$
in Eq. \eqref{eq:Averaging} lie in a compact interval within $(0,1)$,
such that a true averaging takes places when interactions happen.
Formally, the phases of neurons can achieve values $\phi_{i}\in(-\infty,\phi_{\Theta,i}]$.
Each neuron fires at least once within a time interval of length $T$,
therefore the phases are certainly bounded to the compact interval
$\phi_{i}\in[-T+\phi_{\Theta,i},\phi_{\Theta,i}]$. Further, in inhibitory
networks the phase after an interaction is smaller than before, \begin{equation}
H_{\eps}^{(i)}(\phi)=U_{i}^{-1}(U_{i}(\phi)+\eps)<U_{i}^{-1}(U_{i}(\phi))=\phi,\end{equation}
because together with $U_{i}$ also $U_{i}^{-1}$ is strictly monotonic
increasing and $\eps<0$. The strict concavity of $U_{i}\left(\phi\right)$
implies \begin{equation}
0<\frac{\partial H_{\eps}^{(i)}\left(\phi\right)}{\partial\phi}=\frac{U_{i}^{\prime}\left(\phi\right)}{U_{i}^{\prime}\left(H_{\eps}^{(i)}\left(\phi\right)\right)}<1\label{eq:boundingofHprime}\end{equation}
for any finite $\phi$ (cf.~Fig.~\ref{fig:visualH} for illustration).
The derivative $\frac{\partial H_{\eps}^{(i)}\left(\phi\right)}{\partial\phi}$
is continuous in $\phi$, therefore the image of $[-T+\phi_{\Theta,i},\phi_{\Theta,i}]$
under the map $\frac{\partial H_{\eps_{ij}}^{(i)}\left(\phi\right)}{\partial\phi}$
is compact. Together with Eq.~\eqref{eq:boundingofHprime} it follows
that \begin{equation}
0<\cmn,\cmx<1.\label{eq:BoundsOfC}\end{equation}

Taken together, Eq.~(\ref{eqn:UpdateDeltaPhiOnSending}, \ref{eqn:UpdateDeltaSigmaOnSending},
\ref{eq:ShiftOfnewSpike}, \ref{eq:Averaging}, \ref{eq:BoundsOfC})
imply that a true averaging between shifts already present in the
system takes place when a spike is received. For other events the
shifts stay unchanged. As a consequence, the maximal and minimal shift
after the $n$th event,\begin{equation}
\Dmx^{(n)}:=\max_{i,j,k}\left\{ \Delta_{i}^{(n)},\Delta\sigma_{ijk}^{(n)}\right\} \quad\txt{and}\quad\Dmn^{(n)}:=\min_{i,j,k}\left\{ \Delta_{i}^{(n)},\Delta\sigma_{ijk}^{(n)}\right\} ,\label{eq:DefDmxAndDmn}\end{equation}
are bounded by the initial shifts for all future events, \begin{equation}
\Dmx^{(n)}\leq\Dmx^{(0)}\quad\txt{and}\quad\Dmn^{(n)}\geq\Dmn^{(0)},\label{eq:DeltaBound}\end{equation}
as long as the order of events in both sequences is the same. Here
the minima and maxima are taken over $i,j\in\left\{ 1,\ldots,N\right\} $
and $k$ numbers the spikes traveling from neuron $j$ to $i$ at
time $t_{n}$. An initial perturbation cannot grow, thus the trajectory
is Lyapunov stable. We note that we did not make any assumptions about
the network connectivity, the results hold for any network structure
and the described class of trajectories.

\subsection{Asymptotic stability\label{sec:asymptotic-stability}}

In this section we prove that for strongly connected networks even
asymptotic stability holds under the condition that the perturbed
and the unperturbed sequences have the same order of events, i.e.
the order of events is unchanged by small perturbations. The central
idea is as follows: We study the dynamics and convergence of two neighboring
trajectories. We will track the propagation of the perturbation of
one specific neuron $l_{0}$ through the entire network. Since there
is a directed connection between every pair of neurons in the network
and any spike reception leads to an averaging of shifts, there is
an averaging over all perturbations in the network. For large times
all perturbations converge towards the same value, such that both
sequences become equivalent, only shifted by a constant temporal offset.
Further details and the derivation of the following Eqs.~\eqref{eq:BoundsofDmxK},\eqref{eq:BoundsOfDmnK}
and \eqref{eq:BoundsCAst} is provided in the appendix.

We find the upper bound

\begin{equation}
\Dmx^{(K)}\leq c^{\ast}\cdot\Delta_{l_{0}}^{(0)}+\left[1-c^{\ast}\right]\cdot\Dmx^{(0)}\label{eq:BoundsofDmxK}\end{equation}
for the maximal perturbation after $K:=2NM$ events and analogously
the lower bound 

\begin{equation}
\Dmn^{(K)}\geq c^{\ast}\cdot\Delta_{l_{0}}^{(0)}+\left[1-c^{\ast}\right]\cdot\Dmn^{(0)}\label{eq:BoundsOfDmnK}\end{equation}
for the minimal perturbation. The averaging factor $c^{\ast}$ is
determined by the network parameters and bounded to\begin{equation}
3/4\leq1-c^{\ast}<1.\label{eq:BoundsCAst}\end{equation}
A bound for the difference of the maximal and minimal perturbation after $K$
events is therefore given by\begin{equation}
\Dmx^{(K)}-\Dmn^{(K)}\leq\left(1-c^{\ast}\right)\left(\Dmx^{(0)}-\Dmn^{(0)}\right).\label{eq:Convergence}\end{equation}

\begin{figure}
\begin{centering}
\includegraphics[clip,width=85mm]{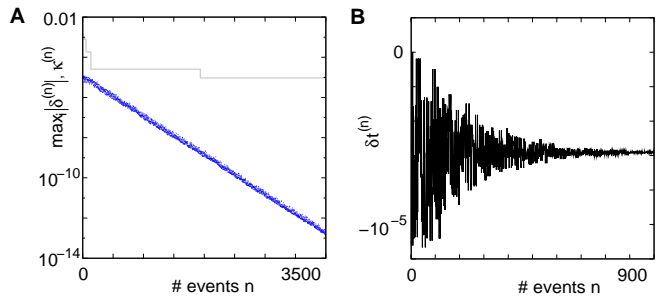}
\par\end{centering}

\caption{Stable dynamics in the network of Fig.~\ref{fig:balanced}(A-C).
(A) Exponential decay of the maximal perturbation $\max_{i}\left|\delta_{i}^{(n)}\right|$
(blue dots) and minimal margin $\kappa^{(n)}$ (gray line) for one
microscopic dynamics. The initial perturbation $\max\left|\delta_{i}^{(0)}\right|\approx10^{-5}$
decays exponentially fast. (B) Exponential convergence of the temporal
offset $\delta t^{(n)}$. For large $n$ the original and the perturbed
sequences are equivalent, just shifted by a constant offset $\delta t=\lim_{n\rightarrow\infty}\left\{ \delta t^{(n)}\right\} .$
\label{fig:convergence}}

\end{figure}

The spread of any perturbation through the network has a contracting
effect on the total perturbation, it leads to a decay of the difference
between the extremal perturbations at least by a factor $(1-c^{\ast})$.
Inequality~\eqref{eq:Convergence} implies together with the bound~\eqref{eq:BoundsCAst}
that for the considered trajectories in the long-time limit the maximal
and minimal perturbation are the same, \begin{equation}
\lim_{n\rightarrow\infty}\Dmx^{(n)}=\lim_{n\rightarrow\infty}\Dmn^{(n)}.\end{equation}
Thus, for $t\rightarrow\infty$ the events are just shifted by a constant
temporal offset \begin{equation}
\delta t=\lim_{n\rightarrow\infty}\delta t^{(n)}=-\lim_{n\rightarrow\infty}\Dmx^{(n)}\end{equation}
(cf.~Fig.~\ref{fig:convergence} for illustration), and both sequences
become equivalent $\lim_{n\rightarrow\infty}\delta_{i}^{(n)}=0$.
Thus all sequences considered are asymptotically stable for all strongly
connected networks and perturbations decay exponentially fast with
at least \begin{equation}
\Dmx^{(n)}-\Dmn^{(n)}\leq\left(1-c^{\ast}\right)^{\left\lfloor \frac{n}{K}\right\rfloor }\left(\Dmx^{(0)}-\Dmn^{(0)}\right),\label{eq:ExpnConvergence}\end{equation}
where $\left\lfloor .\right\rfloor $ is the floor function. The actual
and numerically measured exponential decay is much faster than the
estimation given by Eq.~\eqref{eq:ExpnConvergence}, because for deriving Eqs.~\eqref{eq:BoundsofDmxK} and \eqref{eq:BoundsOfDmnK} in the appendix we had to assume a worst-case scenario. 
The main reasons for the faster decay
are that (i) the mean path length is more meaningful for estimating
the number of events until all neurons have received an input from
the starting one and (ii) it is impossible that the neurons receive
the worst-case perturbation at each reception.

\subsection{Margins}

The stability results in the previous sections hold for the class
of patterns, where a small perturbation does not change the order
of spikes. In this section we show that typical spike patterns, generated
by networks with a complex connectivity, belong to this class. 

In heterogeneous networks with purely inhibitory interactions the
occurrence of events at \emph{identical} times has a zero probability.
There is no mechanism causing simultaneous spiking, like supra-threshold
inputs in excitatorily coupled networks. As long as two events do
not occur at the same event time, there is a non-zero perturbation
size keeping the order unchanged in any \emph{finite} time interval.
However, the requirement of an unchanged event order yields more and
more conditions over time such that the allowed size of a perturbation
could decay more quickly with time than the actual perturbation. This
is excluded if a temporal \emph{margin} $\mu^{(n)}$ (cf.~\citep{Jin2002})
stays larger than the dynamical perturbation for \emph{infinite} time.
Formally, after time $t_{n}$ denote the $k$th potential future event
time (of the original trajectory) that would arise if there were no
future interactions by $\theta_{n,k}$, $k\in\mathbb{N}$, and the
temporal \emph{margin} by $\mu^{(n)}:=\theta_{n,2}-\theta_{n,1}$.
A sufficiently small perturbation, satisfying $\Dmx^{(n)}-\Dmn^{(n)}<\mu^{(n)}$,
cannot change the order of the $(n+1)$th event.

\textbf{Stability of generic periodic orbits. }This directly implies
that almost all \emph{periodic orbits} (all those with non-degenerate
event times $t_{n}$) consisting of a finite number of $P$ events
are stable because there is a minimal margin \begin{equation}
\kappa^{(P)}:=\min\limits _{n\in\{1,\ldots,P\}}\mu^{(n)}\label{eq:defMinMargin}\end{equation}
 for every non-degenerate periodic pattern (c.f.~also \citep{Memmesheimer2009}).

\begin{figure*}
\begin{centering}
\includegraphics[clip,width=140mm]{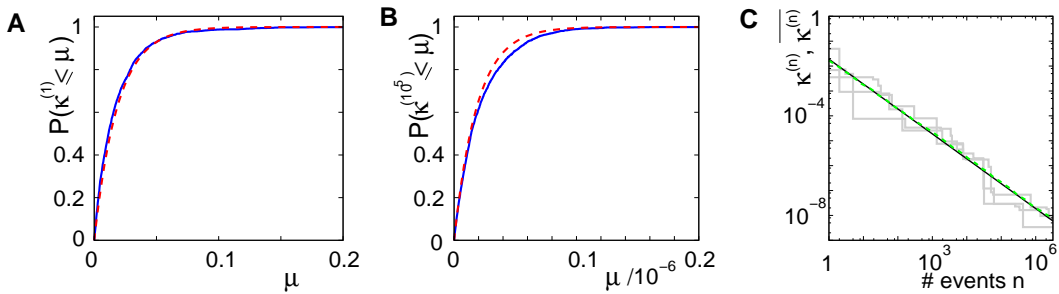}
\par\end{centering}

\caption{Margins in the network given in Fig.~\ref{fig:balanced}(A-C). (A,B)
Probability distribution $P(\kappa^{(n)}\leq\mu)$ after (A) $n=1$
and (B) $n=10^{5}$ events. The blue curve is measured distribution
over $2500$ samples, the red dotted line is the analytical prediction
(no free fit parameter, rate $\nu$ is measured; cf. Eq. \eqref{eq:minMarginDistr}).
(C) Algebraic decay of the average minimal margin, $\overline{\kappa^{(n)}}$
(green dashed line, averaged over $250$ random initial conditions)
and its analytical prediction (no free fit parameter; black solid
line). Additionally we show the minimal margin $\kappa^{(n)}$ for
three exemplary initial conditions (gray lines), including that of
Fig.~\ref{fig:convergence}(A). \label{fig:margins}}

\end{figure*}

\textbf{Stability of irregular non-periodic spiking activity. }We
study the stability properties of \emph{irregular non-periodic spike
sequences} by considering the minimal margin $\kappa^{(n)}$ over
the first $n$ events. For simplicity, we consider delay distributions
where $\tau_{ij}$ is independent of the spike receiving neuron $i$.
The irregular spiking dynamics of the entire network is well modeled
by a Poisson point process with rate $\nu^{\tsf}$, where $\nu^{\tsf}$
specifies the mean firing rate of the network \citep{Brunel1999,Brunel2000,Tuckwell1988,Burkitt2006}.
We assume that, along with the irregular dynamics, the temporal margins
are also generated by a Poisson point process with the network event
rate $\nu=2\nu^{\tsf}$, because any spike sending time generates
one receiving event due to the independence of $\tau_{ij}$ from $i$
and the definition of events. The distribution function of margins
is given by\begin{equation}
P\left(\mu^{(n)}\leq\mu\right)=1-e^{-\nu\mu}.\end{equation}
 Therefore, the probability that the minimal margin $\kappa^{(n)}$
after $n$ events is smaller or equal to $\mu$ is determined by the
probabilities that not all individual margins $\mu^{(n)}$ are larger
than $\mu$ such that \begin{equation}
P\left(\kappa^{(n)}\leq\mu\right)=1-\prod_{m=1}^{n}P\left(\mu^{(m)}>\mu\right)=1-e^{-n\nu\mu}\label{eq:minMarginDistr}\end{equation}
 with density $\rho_{n}(\mu):=dP\left(\kappa^{(n)}\leq\mu\right)/d\mu=n\nu\exp(-n\nu\mu).$
This implies an \emph{algebraic} decay with the number $n$ of events
for the expected minimal margin \begin{equation}
\overline{\kappa^{(n)}}=\int_{0}^{\infty}\mu\rho_{n}(\mu)d\mu=(\nu n)^{-1}\label{eqn:powerlaw}\end{equation}
 that depends only on the event rate and is independent of the specific
network parameters. Numerical simulations show excellent agreement
with this algebraic decay (\ref{eqn:powerlaw}); a typical example
is shown in Fig.~\ref{fig:margins}(C).

This already strongly suggests that a sufficiently small perturbation
stays smaller than the minimal margin for all times. However, in each
step, the exponential distribution of margins has finite density for
arbitrary small values of $\mu$, i.e. in each step the margin can
fall below the level of perturbation with finite probability. We will
show that $P\left(\exists n\in\mathbb{N}:\mu^{(n)}\leq\Dmx^{(n)}-\Dmn^{(n)}\right)$,
the probability that there is at least one step in which the margin
falls below the perturbation size, goes to zero if the size of the
initial perturbation goes to zero. Thus, also the probability that
there is a change in the order of events goes to zero. Of course,
we cannot expect to reach zero for nonzero perturbation.

We derive a lower bound for the probability that the margin stays
larger than $\Dmx^{(n)}-\Dmn^{(n)}$ for infinite time. We show that
it converges to one when the size of the perturbation goes to zero
and thus prove that sufficiently small perturbations have arbitrarily
high probability to stay smaller than the minimal margin for all times.
We start from the upper bound for the evolution of the perturbation,
Eq.~\eqref{eq:ExpnConvergence}. Using $\left\lfloor \frac{n}{K}\right\rfloor \geq\frac{n}{K}-1$
and Eq.~\eqref{eq:BoundsCAst} leads to\begin{align}
\Dmx^{(n)}-\Dmn^{(n)} & \leq\left(1-c^{\ast}\right)^{\frac{n}{K}-1}\left(\Dmx^{(0)}-\Dmn^{(0)}\right),\nonumber \\
 & =\exp\left(\frac{\log(1-c^{\ast})}{K}n\right)\frac{\Dmx^{(0)}-\Dmn^{(0)}}{1-c^{\ast}},\nonumber \\
 & =C\exp(-\alpha n),\label{eq:Cexpalphan}\end{align}

where we introduced\begin{align}
C:= & \frac{\Dmx^{(0)}-\Dmn^{(0)}}{1-c^{\ast}},\label{eq:DefOfBC}\\
\alpha:= & -\frac{\log(1-c^{\ast})}{K}>0.\label{eq:DefOfAlpha}\end{align}
In particular, $C\rightarrow0$ if the initial perturbation goes to
zero, i.e.~$\Dmx^{(0)}-\Dmn^{(0)}\rightarrow0$, while $\alpha$
is independent of the initial perturbation. The probability, that
all margins are larger than all perturbations is given by\begin{equation}
P\left(\forall n:\mu^{(n)}>\Dmx^{(n)}-\Dmn^{(n)}\right)=\prod_{n=1}^{\infty}P\left(\mu^{(n)}>\Dmx^{(n)}-\Dmn^{(n)}\right),\end{equation}
since the margins are independent. Using Eq.~\eqref{eq:Cexpalphan}
and $P\left(\mu^{(n)}>\mu\right)=\exp(-\nu\mu)$ yields\begin{align}
\prod_{n=1}^{\infty}P\left(\mu^{(n)}>\Dmx^{(n)}-\Dmn^{(n)}\right) & \geq\prod_{n=1}^{\infty}P\left(\mu^{(n)}>C\exp(-\alpha n)\right)\nonumber \\
 & =\prod_{n=1}^{\infty}\exp(-\nu C\exp(-\alpha n))\nonumber \\
 & =\exp\left(-\nu C\sum_{n=1}^{\infty}\exp(-\alpha n)\right)\nonumber \\
 & =\exp\left(-\nu C\frac{1}{\exp(\alpha)-1}\right),\label{eq:expprob}\end{align}
which goes to one if the initial perturbation (and thus $C$) goes
to zero.

We note that the assumption of a constant lower bound of the minimal
margin is not necessary in contrast to \citep{Jin2002}. Indeed this
assumption would be highly problematic, because in an irregular dynamics
arbitrarily small margins naturally occur with positive probability
in every step. So, assuming some lower bound would exclude generic
irregular dynamics. In contrast, the novel approach introduced above
enabled us to show that the generic irregular dynamics is stable.

For infinitely large networks in a mean field approach, where one takes the limit of infinitely many neurons in the beginning (e.g. \cite{Vreeswijk1996, Vreeswijk1998, Brunel1999}), our method is not applicable in a straightforward way. In this limit the average inter-spike interval goes to zero and a positive margin cannot be presumed. In our analysis we employ the fact that the minimal margin stays finite, i.e. a sufficiently small perturbation does not change the order of events. This assumption and therefore our results hold for arbitrarily large finite systems. 

\subsection{Convergence to periodic orbits}

Interestingly, the asymptotic stability together with the finite phase
space imply that generic spike sequences converge to a periodic orbit.
To show convergence analytically, we extend and explicate the ideas
presented in \citep{Jin2002,Jahnke2008}. In the following we focus
on a finite subsequence $s^{\ast}$ of a spike sequence generated
by a given network. The number $E^{\ast}$ of events in $s^{\ast}$
is called the length $E^{\ast}$ of $s^{\ast}.$ As discussed towards
the end of section \ref{sec:Network-model}, the considered system
is finite dimensional with dimension bounded from above by $N+ND'$.
Thus, if $s^{\ast}$ is sufficiently long it contains at least two
disjoint subsequences $s_{1}$ and $s_{2}$ of length $E$, where
the ordering of events is identical. The maximal $E$ for which the
existence of $s_{1}$ and $s_{2}$ is guaranteed, is given by the
largest integer $E$ satisfying \begin{equation}
E^{\ast}\geq\left(N+ND^{\prime}\right)^{E}+E.\end{equation}
When increasing the observation length $E^{\ast}$ also the possible
length $E$ of the subsequences increases. Both the phases and the
variables encoding the spikes in transit are bounded to a finite interval,
$\phi_{i}(t)\in\left[-T+\phi_{\Theta,i},\phi_{\Theta,i}\right]$ and
$\sigma_{ijk}(t)\in\left[0,\max_{ij}\left\{ \tau_{ij}\right\} \right]$
at any given time $t$. Therefore the maximal event-based distance
between two trajectories is also bounded to a finite size,\begin{equation}
\Phi_{\txt{max}}:=\max_{i,j}\left\{ T,\tau_{ij}\right\} .\end{equation}
Thus comparing sequences $s_{1}$ and $s_{2}$ the initial event-based
distance between their underlying trajectories at the beginning of
$s_{1}$ and at the beginning of $s_{2}$ is bounded by $\Phi_{\txt{max}}$,
i.e. $\Dmx^{(0)}-\Dmn^{(0)}\leq\Phi_{\txt{max}}$. By definition,
the order of events in both sequences $s_{1}$ and $s_{2}$ is the
same; therefore the distance between them shrinks according to Eq.~\eqref{eq:ExpnConvergence}.
After $E$ events the distance is bounded by (cf.~Eq.~\eqref{eq:Cexpalphan})
\begin{equation}
\Dmx^{(E)}-\Dmn^{(E)}\leq\Phi_{\txt{max}}^{\ast}:=\frac{\Phi_{\txt{max}}}{1-c^{\ast}}\exp\left(-\alpha E\right),\end{equation}
where we used the definition \eqref{eq:DefOfAlpha}. We note that,
if we increase our observation length $E^{\ast}$ and therefore the
subsequence length $E$, the maximal possible distance $\Phi_{\txt{max}}^{\ast}$
between the trajectories underlying the sequences $s_{1}$ and $s_{2}$
decreases. If $\Phi_{\txt{max}}^{\ast}$ is sufficiently small, i.e.
the distance between $s_{1}$ and $s_{2}$ after $E$ events is smaller
than the average margin, there is a high probability that also the
order of events in the sequence following $s_{1}$ and $s_{2}$ are
the same. Analogous to Eq.~\eqref{eq:expprob} the probability is
given by \begin{equation}
P_{E^{\ast}}=\exp\left(-\nu\frac{1}{\left(1-c^{\ast}\right)\left(\exp(\alpha)-1\right)}\Phi_{\txt{max}}^{\ast}\right),\end{equation}
which goes to one when the observation length $E^{\ast}$ tends to
infinity.

\begin{figure*}
\begin{centering}
\includegraphics[clip,width=140mm]{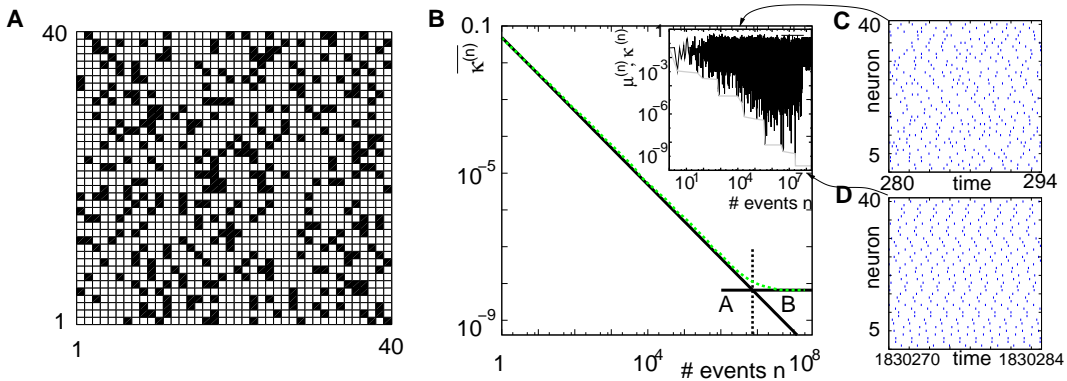}
\par\end{centering}

\caption{Convergence towards a periodic orbit in a random network ($N=40$,
$\gamma_{i}\equiv1$, $I_{i}\equiv3.0$, $V_{\Theta,i}\equiv1.0$,
$\tau_{ij}\equiv T_{i}^{\free}/10$, $\left|\Pre(i)\right|\equiv8$,
$\sum_{j}\epsilon_{ij}\equiv-3.3$). (A) Coupling matrix, each realized
connection is indicated by a black square. (B) Average minimal margin
$\overline{\kappa^{(n)}}$ (averaged over $250$ random initial conditions,
cf. Fig.~\ref{fig:margins}(C)) decays as a power-law (region $A$)
and saturates after about $10^{7}$ events (region $B$) when the
periodic orbit is reached. Inset: Margin $\mu^{(n)}$ (black) and
minimal margin $\kappa^{(n)}$ (gray) for a trajectory started from
one specific initial condition. The margin $\mu^{(n)}$ fluctuates
strongly on the transient and is comparatively large after the sequence
becomes periodic; thus the minimal margin $\kappa^{(n)}$ does not
decrease further for future events $n$. (C), (D): Snapshots of irregular
spike sequences (C) after $n\approx15.000$ events on the transient
and (D) after $n\approx10^{8}$ events on the periodic orbit. \label{fig:periodic}}

\end{figure*}

This implies a periodic dynamics: Assume that $s_{1}$ occurs first
in $s^{\ast}$ at the $a$th event and after a certain event number
$a+L\leq E^{\ast}-E$ the sequence $s_{2}$ begins. We have seen that
(with a certain probability) the ordering of events for (infinite)
sequences starting at the $a$th and the $(a+L)$th event is the same
for all future times. Therefore also the sequence starting at the
$(a+2L)$th event is identical to the ones mentioned before, so the
ordering of events is periodic. Together with the exponential convergence
of equally ordered sequences this implies that also the spike timings
converge towards a periodic orbit. For arbitrarily long observed sequences
this happens with an arbitrary large probability that tends to one
as $E^{\ast}\rightarrow\infty$.

\begin{figure*}
\begin{centering}
\includegraphics[clip,width=140mm]{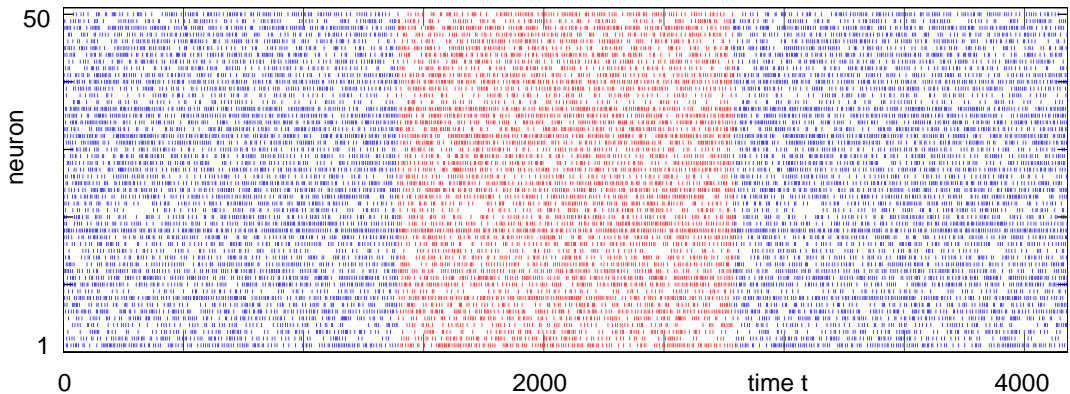}
\par\end{centering}

\caption{Long periodic orbit in a random network ($N=50$, $\gamma_{i}\equiv1$,
$I_{i}\equiv1.2$, $V_{\Theta,i}\equiv1.0$, $\tau_{i,j}\equiv0.1T_{i}^{\free}$).
The coupling strength are randomly independently drawn from the uniform
distribution $\left[-1,0\right]$ and normalized afterwards ($\sum_{j=1}^{N}\eps_{ij}=-6.5$).
The final spike pattern (shown after a transient of $9\cdot10^{6}$
events) repeats every $11012$ events and is highly irregular. \label{fig:periodicII}}

\end{figure*}

In Fig.~\ref{fig:periodic} we show a typical example: The mean margin,
$\overline{\kappa^{(n)}}$, decays algebraically on the transient
and saturates after the periodic orbit is reached. Interestingly,
the periodic attractor (shown in Fig.~\ref{fig:periodic}(D)) of
the sparse random network resembles the ''splay state'' known from
globally coupled networks \citep{Strogatz1993}. In a splay state
the firing pattern is characterized by equally spaced inter-spike-intervals.
It has been shown that it is possible to design networks, which exhibit
more complex periodic spike patterns \citep{Memmesheimer2006,Memmesheimer2006a}.
Indeed, in different parameter regimes we observe such complex periodic
orbits, with a large periodicity, cf.~the heterogeneous globally
coupled network in Fig.~\ref{fig:periodicII}, where the periodic
orbit is reached after a small number of events compared to sparse
networks. As shown previously \citep{Timme2002b,Timme2008}, highly
irregular spiking activity may coexist with even the simplest (fully
synchronous) periodic orbits that exhibits regular, maximally ordered
activity.

\begin{figure}
\begin{centering}
\includegraphics[clip,width=170mm]{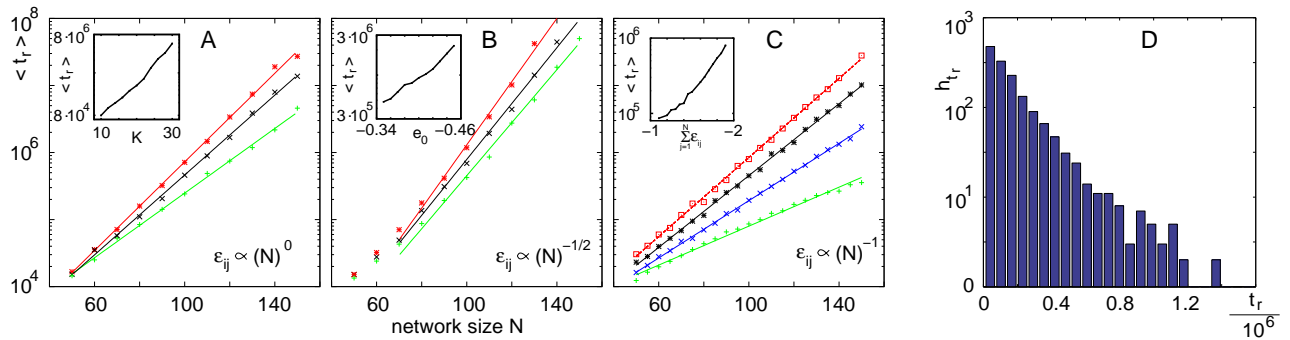}
\par\end{centering}

\caption{Scaling of the transient lengths in sparse random networks.
(A) The sum of the external excitation and the internal inhibition is fixed, $\hat{I_i}:=\sum_{j=1}^{N} \eps_{ij} + I_i \equiv 1.0$, such that the mean firing rate, $\left<\nu_{i}\right>_{i}$,
of the single neurons stays approximately constant. The number of incoming connections to each neurons, $K_i:=\left|\mbox{Pre}(i)\right|$, is fixed to $K_i\equiv15$ (green~$+$), $K_i\equiv20$ (black~$\times$), $K_i\equiv25$ (red~$\ast$). The single connections are set to $\eps_{ij}=e_0/\sqrt{K_i}$ with $e_0=-0.38$. The solid lines show the best exponential fit. We observe an exponential increase of the trial averaged transient length $\left<t_r \right>$ with network size $N$. The inset shows the dependence of the transient length on the in-degree for fixed network size ($N=100$). 
(B) Networks with fixed fraction of connections, where $K_i\equiv\left\lfloor pN\right\rfloor$ and $p=1/4$. We scale the coupling strengths as $\sum_{j=1}^N\eps_{ij} = e_0\sqrt{K_i}$, such that $\eps_{ij}\propto 1/\sqrt{K_i}$, and the external current $I_i$ is adjusted appropriately to fix the mean firing rate $\left<\nu_i\right>_i$. The mean transient lengths for different $e_0$ are shown together with the best exponential fit ($e_0=-0.34$ (green~$+$), $e_0=-0.38$ (black~$\times$), $e_0=-0.42$ (red~$\ast$)). Again, we observe an exponential increase of the transient lengths with $N$. The inset shows the increase of the transient length with inhibitory coupling strength for fixed network size ($N=100$).
(C) Networks with fixed fraction of connections, where $K_i\equiv\left\lfloor pN\right\rfloor$ and $p=1/4$. The external currents $I_{i}\equiv3.0$ are fixed and the internal coupling is normalized to $\sum_{j=1}^{N}\eps_{ij}\equiv-1.1$ (green~
$+$), $-1.5$ (blue~$\times$), $-1.75$ (black~$\ast$), $-1.9$
(red~$\square$), such that the mean firing rate, $\left<\nu_{i}\right>_{i}$,
of the single neurons is approximately the same for each curve. The number of incoming connections per neuron, $K_i$, increases linearly with network size, so the coupling strengths are scaled as $\eps_{ij}\propto 1/K_i$.
Here we also observe an exponential increase of  $\left<t_{r}\right>$  with network size. The inset shows the fast increase
of the transient length with inhibitory coupling strength for fixed network size and external current ($N=100, I_{i}\equiv3.0$). 
(D) The transient length is broadly distributed
as shown in the histogram for $2500$ trials started from random initial
conditions, where the initial phases were randomly independently drawn from the uniform distribution on $[0,1]$ ($N=100,$ $I_{i}\equiv3.0,$ $\sum_{j=1}^{N}\eps_{ij}\equiv-1.5$)
\label{fig:transient}
}
\end{figure}

Although the attractor is reached after a finite number of events,
the transient becomes very important in systems with strong inhibition
or large number of neurons. As formerly found in networks with excitatory
coupling \citep{Zumdieck2004}, and also in weakly diluted networks
with purely inhibitory interactions \citep{Zillmer2006}, the transient
length grows rapidly with network size such that the dynamics is governed
by the transient for large time scales. We study inhibitory random
networks with an arbitrary network structure, typically far away from
the weakly diluted topology. To perform numerical measurements of
the transient length in dependence on the network size $N$, we define
the length of the transient, $t_{r}$ by the number of events after
which the order of events stays periodic. When increasing the network
size $N$, we leave the sum $\hat{I}:=I_i +\sum_{j=1}^N \eps_{ij}$ of the external excitatiory current, $I_i$, and the internal inhibition, $\sum_{j=1}^{N} \eps_{ij}$, constant. Thus, on average each neuron receives a constant effective input independent
of $N$ and the mean firing rate of a single neuron $\left<\nu_{i}\right>$ is
approximately conserved. Fig.~\ref{fig:transient} shows the increase
of transient lengths with network size for different sizes of internal inhibition and different scalings of the single connection strengths.
We observe an exponential increase of the mean transient length with network size $N$ independent of the scaling of the coupling strengths, $\eps_{ij}$, and the number of incoming connections, $K_i:=\left|\mbox{Pre}(i)\right|$. This is qualitatively similar to the scaling of the transient lengths in weakly diluted networks  \citep{Zillmer2006}. Assuming that the rapid increase continues for much larger networks, the transient will dominate the dynamics essentially forever in networks of biological relevant sizes. In this sense the transient becomes quasi-stationary (cf. also \cite{Zillmer2006, Zillmer2009}). If a larger network is in the balanced state (cf.~Fig.~\ref{fig:balanced}), the stable transients typically dominate the network dynamics. 

\subsection{Robustness of stability and smooth transition to chaos}

In the following we will check the robustness of our results. The
considerations above hold for networks with inhibitory interactions
without temporal extent. We investigate the influence of excitatory
interaction and pulses with a finite duration. For small deviation
from the networks considered above the stability properties are similar,
for large fractions of excitatory neurons and large temporal extent
we observe a transition to a chaotic regime. For temporally extended
couplings we assume that after a neuron is reset all previous input
is lost. Therefore the state of a neuron is specified by its last
spike time and all spikes it has received afterwards. The phase representation
is thus not meaningful anymore and we track two trajectories by comparing
the differences in the last spiking times of the neurons and in the
spike arrival times since these last spikings. To keep the section
consistent, we adopt this view when studying the Lyapunov exponents
of the excitatory dynamics.

\subsubsection*{Excitatory interactions}

\begin{figure*}
\begin{centering}
\includegraphics[clip,width=170mm]{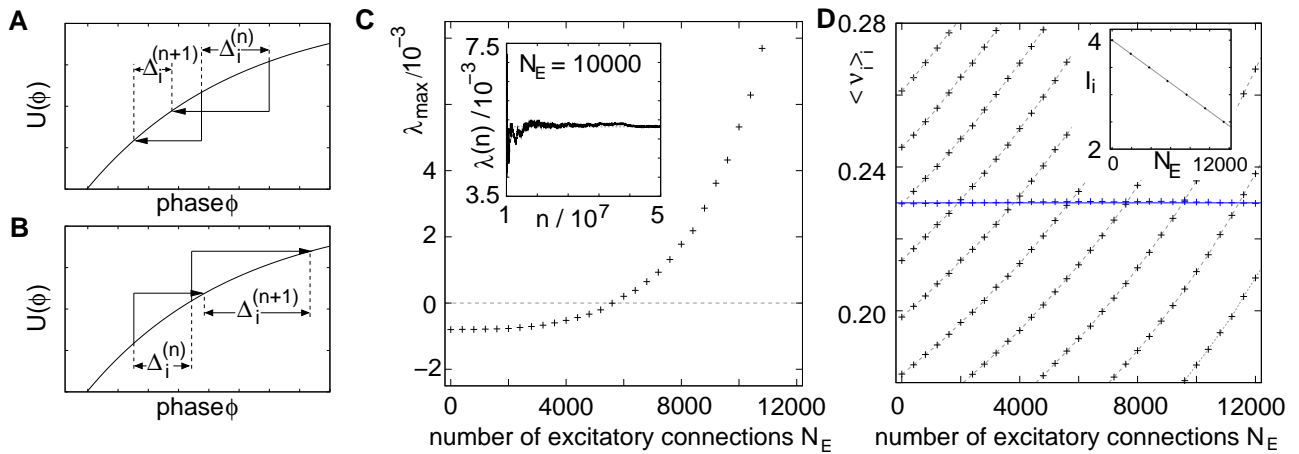}
\par\end{centering}

\caption{Destabilizing effect of excitation. (A) Simultaneous receiving of
an inhibitory input decreases the phase shift. (B) Simultaneous receiving
of an excitatory input increases the phase shift. (C) The largest
Lyapunov exponent is measured for an increasing fraction of excitatory
neurons starting with the network of Fig.~\ref{fig:balanced}(A-C)
and ending with the network of Fig.~\ref{fig:balanced}(D-F). Here
we keep the firing rate constant. Inset shows exemplarily the convergence
of a finite time Lyapunov exponent with $n$. (D) Average firing rate
$\left<\nu_{i}\right>_{i}$ versus number of excitatory connections in the network
for different external currents $I_{i}$ (black crosses, values belonging
to the same $I_{i}$ are connected by a dashed line). The neurons'
firing rate stays almost constant, if we reduce the external current
linearly with the number of excitatory neurons (blue crosses). The
inset displays the current strength employed to maintain firing rate
of $\left<\nu_{i}\right>_{i}\approx0.23$. (Further details see text.) \label{fig:exitat}}

\end{figure*}

We have shown that in networks with purely inhibitory interactions
the dynamics is typically stable. If the connection from neuron $j$
to neuron $i$ is excitatory, the phase shift $\Delta_{i}^{(n+1)}$
of the postsynaptic neuron $i$ after receiving a spike from neuron
$j$ as the $(n+1)$th event may exceed its shift $\Delta_{i}^{(n)}$
before and the shift $\Delta\sigma_{ij1}^{(n)}$ of the received spike.
Fig.~\ref{fig:exitat}(A,B) gives an illustration: A spike is simultaneously
received in the perturbed and in the unperturbed dynamics (i.e. $\Delta\sigma_{ij1}^{(n)}=0$).
The phase shift before and after the application of the transfer function
$H_{\eps}^{(i)}(\phi)$ is shown for (A) an inhibitory input and (B)
an excitatory input. For inhibitory input the phase shift $\Delta_{i}^{(n)}$
is reduced, this leads to the stable dynamics as described in the
sections above. For excitatory input, the phase shift $\Delta{}_{i}^{(n)}$
increases when the spike is received. 

Indeed, since the inverse of $U_{i}(\phi)$ is monotonically increasing
with $\phi,$ we find for a given $\eps>0$ \begin{equation}
H_{\eps}^{(i)}(\phi)=U_{i}^{-1}(U(\phi)+\eps)>U_{i}^{-1}(U_{i}(\phi))=\phi.\end{equation}
In contrast to Eq.~\eqref{eq:boundingofHprime} the derivative of
the transfer function $H_{\eps}^{(i)}(\phi)$ is bounded from below
by\begin{equation}
\frac{dH_{\eps}^{(i)}\left(\phi\right)}{d\phi}=\frac{U_{i}^{\prime}\left(\phi\right)}{U_{i}^{\prime}\left(H_{\eps}^{(i)}\left(\phi\right)\right)}>1.\end{equation}
According to Eq.~\eqref{eq:Averaging}, this can lead to an increase
of (in particular extremal) perturbations and to a destabilization
of the trajectory. The upper bound of $c_{ij}^{(n)}$, $\cmx<1$ (cf.
Eq.~\eqref{eq:BoundsOfC}) does not hold anymore. However, in a network
with a small fraction of excitatory connections, the trajectory is
still stable. At an interaction the perturbation may increase, but
the stabilizing effect of inhibitory inputs dominates the dynamics.
We study the transition from the stable regime to chaotic dynamics
(a discussion of the chaotic dynamics in networks with purely excitatory
interactions can be found in \citep{Zumdieck2004}). When increasing
the number of excitatory couplings, we increase the mean effective
input current to the neurons. Thus we additionally decrease the external
input $I_{i}$ to keep the network rate $\nu$ constant. Indeed, in
good approximation, the current has to be decreased linearly with
$N_{E}$, the number of excitatory connections, $I_{i}\equiv I-kN_{E}$
where $I$ is the original input current. 

To quantify the transition we estimate the largest Lyapunov exponent
of the system: At the $n$th event, we denote $n-W(n)$ the earliest
event which still influences the future dynamics of the system explicitly.
We apply an initial perturbation of size $\left\Vert \Delta_{0}\right\Vert _{1}$
to the event times $t_{0},t_{-1},\ldots,t_{-W(0)}$, where $\Delta_{n}=(\Delta t_{n},\Delta t_{n-1},\ldots,\Delta t_{n-W(n)})$
is the perturbation vector at the $n$th event time and $\Delta t_{i}$
is the perturbation of $t_{i}$. We evolve the system and rescale
the perturbation vector $\Delta_{n}$ by $a_{n}$ after each event,
such that the rescaled perturbation vector is of the same size as
the initial perturbation, 
\begin{equation}
\left\Vert \Delta_{n}^{\prime}\right\Vert _{1}=\left\Vert \Delta_{n}\cdot a_{n}\right\Vert _{1}=\left\Vert \Delta_{0}\right\Vert _{1}. \label{eq:Drescale}
\end{equation}

The largest Lyapunov exponent, $\lambda_{\txt{max}}$, is then given
by\begin{equation}
\lambda_{\txt{max}}=\lim_{n\rightarrow\infty}\lambda(n)\quad\txt{with}\quad\lambda(n):=\frac{1}{n}\sum_{i=1}^{n}\ln(a_{i}^{-1}).\end{equation}
We observe a transition from a stable to a chaotic regime, characterized
by a positive Lyapunov exponent. For small fraction of excitatory
neurons the dynamics is typically stable, the effect of the inhibitory
pulses dominates the dynamics and, on average, a perturbation do not
grow over time. With increasing $N_{E}$ the Lyapunov exponent increases
until the dynamic becomes chaotic. Of course, in our simulations we
can only study finite time Lyapunov exponents with very large $n$
and estimate the value to which they converge. The chaotic dynamics may thus be transient. However, it dominates the dynamics at least over very long times. 

Estimating the maximal Lyapunov exponent in networks including excitatory interactions can be difficult \citep {Hansel1998, Zumdieck2004, Brette2007, Cessac2008, Kirst2009}. Suprathreshold excitation, together with the infinitely fast response of neurons receiving a spike and the sharp threshold may induce synchronous events. Thus even an infinitesimal small perturbation may change the order of spikes.
Nonetheless generically the perturbation will stay infinitesimal small, in particular for a small fraction of excitatory connections, such that we estimate the largest Lyapunov exponent in the following way: 
We evaluate at each time step the resulting temporal perturbation on the actual event as a result of earlier perturbation under the assumption that the order of spikes stays the same. This gives us the new perturbation vector $\Delta_n$, which is rescaled according to \eqref{eq:Drescale}.
For long times $\lambda (n)$ then will give an estimate of the largest Lyapunov exponent and describe the generic behavior of the trajectory under the influence of sufficiently small perturbations.
However, we cannot exclude the occurrence of macroscopic perturbations in general.
 
Fig.~\ref{fig:exitat}(C,D) shows some numerical results:
In Fig.~\ref{fig:exitat}(C) the largest Lyapunov exponent is measured for an increasing fraction
of excitatory connections starting with the network of Fig.~\ref{fig:balanced}(A-C)
and ending with the network of Fig.~\ref{fig:balanced}(D-F). The
number of excitatory connections is increased by successively choosing
one incoming inhibitory connection per neuron to be excitatory. The
external current, $I_{i}$, is reduced linearly to keep the network
rate unchanged according to $I_{i}\equiv I-kN_{E}$, where $I=4.0$
is the initial external current, $k\approx0.052$ and $N_{E}$ is
the number of excitatory connections. For a large fraction of excitatory
couplings we observe a transition to an unstable, chaotic regime.
The inset demonstrate the convergence of $\frac{1}{n}\sum_{i=1}^{n}\ln(a_{i}^{-1})\rightarrow\lambda_{\txt{max}}$ (exemplary
shown for $N_{E}=1000$). In panel (D) Average firing rate $\left<\nu_{i}\right>_{i}$.
For a constant external current the rate increases with increasing
fraction of excitatory connections (black crosses, for $I_{i}\in\left\{ 2.25,2.5,2.75,3.0,3.25,3.5,3.75,4.0,4.25,4.5\right\} $)
. The neurons' firing rate stays almost constant, if we reduce the
external current linearly with the number of excitatory neurons (blue
crosses). We determined the value of $I_{i}$ at the intersection
point of the $\left<\nu_{i}\right>_{i}$ vs. $N_{E}$ curves with the desired
frequency by linear interpolation. The values ${(I}_{i},N_{E})$ that
give rise to the desired frequency lie in good approximation on a
straight line with slope $-k\approx-0.052$ (cf.~Inset to (D)).

The result is particularly remarkable since in mean-field descriptions
of balanced networks, as long as the mean input to each cell is the
same, the regime where $N_{E}=0$ is comparable to the regime where
$N_{E}>0$ with appropriately reduced external excitatory current
$I_{i}$. 

\subsubsection*{Temporally extended interactions}

Up to now, we considered $\delta$-coupling, where the response to
an action potential is instantaneous. However, in biological neuronal
systems the postsynaptic current has finite temporal extent. In the
following, we investigate the influence of such temporally extended
interactions. The analysis gets more complicated, because neurons
are permanently influenced by incoming signals. As mentioned above,
in our model we assume that the neuron looses the information about
previously received spikes when it reaches the threshold and is reset.

We modify Eq.~\eqref{eq:Vdot} by introducing a temporally extended
interaction kernel $g(t),$ such that the evolution of the membrane
potential is given by\begin{equation}
\frac{d}{dt}V_{i}=f_{i}(V_{i})+\sum_{j=1}^{N}\sum_{k\in\mathbb{Z}}\eps_{i,j}g\left(t-t_{jk}^{\tsf}-\tau_{ij}\right).\label{eq:Vdotext}\end{equation}
In the following analysis we consider single exponential couplings,
$g(t)=\Theta(t)\cdot\beta e^{-\beta t}$ with time scales $\gamma^{-1}>\beta^{-1}>0$,
the time constant of the postsynaptic current is shorter than the
membrane time constant. As an exemplary neuron model we study the
leaky integrate-and-fire neuron, $f_{i}(V_{i})=-\gamma V_{i}(t)+I_{i}$,
but the analysis can easily be extended to more complex neuron models
and interaction kernels. 

\begin{figure*}
\begin{centering}
\includegraphics[clip,width=140mm]{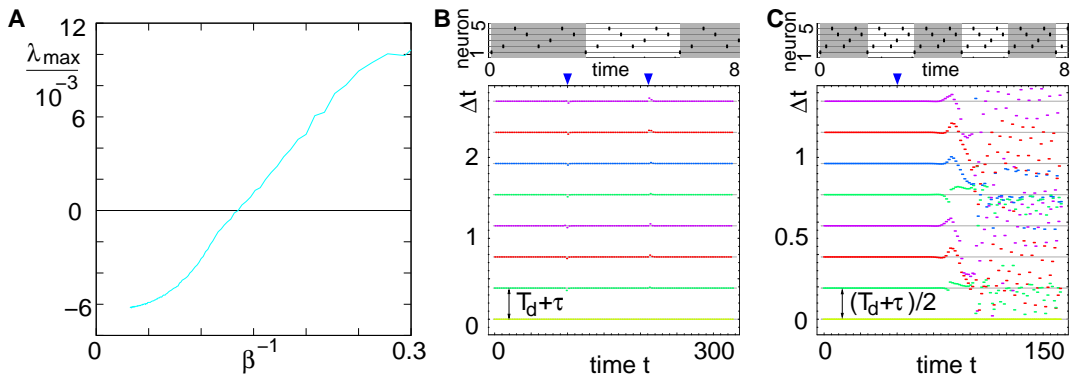}
\par\end{centering}

\caption{Robustness and transition to chaos with increasing temporal extent
of the postsynaptic current. (A) Largest Lyapunov exponent $\lambda_{\txt{max}}$in
a random network ($N=50,$ $|\text{Pre}(i)|\equiv10$, $\gamma=1$,
$I_{i}\equiv4.0$, $\sum_{j}\eps_{ij}\equiv-3.3,$$\tau_{ij}\equiv0.1T_{i}^{\free}$)
versus time constant $\beta^{-1}$ of the synaptic current. (B) Stable
periodic pattern in a network with temporally extended interactions.
(C) Time compressed, thus unstable pattern in a network with temporally
extended interactions. (For details see text.) \label{fig:extended}}

\end{figure*}

Numerical simulations show that the stability of the dynamics is robust
against introduction of synaptic currents with small temporal extent,
but on increase of temporal extension a transition to chaos occurs.
In Fig.~\ref{fig:extended}(A), the largest Lyapunov exponent,
$\lambda_{\txt{max}}$, in a random network is estimated in dependence
of the decay time constant $\beta^{-1}$ of the synaptic current.
For small time constant $\beta^{-1}$, the dynamics behaves similar
to the dynamics with $\delta$-pulse interactions, in particular it
is stable, the largest Lyapunov exponent is negative. For increasing
$\beta^{-1}$ the temporal extension becomes more and more influential
and there is a transition to an unstable, chaotic regime with positive
largest Lyapunov exponent.

We now study the linear stability properties analytically. We denote
the last spiking time of neuron $i$ before $t_{n}$ by \begin{equation}
t_{0}(n,i)=\max_{k\in\mathbb{Z}}\left(\left.t_{ik}^{\tsf}\right|t_{ik}^{\tsf}\leq t_{n}\right);\end{equation}
at $t=t_{0}(n,i)$ the potential of neuron $i$ was reset to zero.
The solution of Eq. \eqref{eq:Vdotext} together with the initial
condition $V_{i}(t_{0}(n,i))=0$ is then given between the $n$th
and $(n+1)$th network event by \begin{eqnarray}
V_{i,n}(t) & = & \frac{I_{i}}{\gamma}\left(1-e^{-\gamma\left(t-t_{0}(n,i)\right)}\right)+\nonumber \\
 &  & \quad\frac{\beta}{\beta-\gamma}\sum_{j=1}^{N}\sum_{k\in\mathbb{Z}}\eps_{ij}\Theta\left(t_{ijk}^{\trf}-t_{0}(n,i)\right)\Theta\left(t-t_{ijk}^{\trf}\right)\left(e^{-\gamma\left(t-t_{ijk}^{\trf}\right)}-e^{-\beta\left(t-t_{ijk}^{\trf}\right)}\right),\label{eq:Vin}\end{eqnarray}
where $t_{ijk}^{\trf}=t_{jk}^{\tsf}+\tau_{ij}$ is the reception time
of the spike sent at $t_{jk}^{\tsf}$ by neuron $j$ at neuron $i$.
The sum in Eq.~\eqref{eq:Vin} takes into account all spikes which
are received by neuron $i$ between $t_{0}(n,i)\leq t\leq t_{n}$
and therefore influence the potential $V_{i,n}(t)$. In the limit
of very short temporal extension of the postsynaptic current, $\beta\rightarrow\infty$,
Eq.~\eqref{eq:Vin} becomes a solution of Eq.~\eqref{eq:Vdot}.
After the $n$th event neuron $i$ would reach the threshold at some time $t^{\prime}$
under the assumption that there are no further inputs after $t_{n}$.
According to Eq.~\eqref{eq:Vin}, $t^{\prime}$ is implicitly given
by \begin{equation}
V_{\Theta,i}-V_{i,n}(A,t=t^{\prime})=0,\label{eq:tprime}\end{equation}
where $A$ is the vector of the original event times $t_{n},\ldots,t_{n-W}$,\begin{equation}
A:=\left(\begin{array}{ccc}
t_{n} & ,\ldots, & t_{n-W}\end{array}\right),\end{equation}
where we introduced $W=\max_{n}\left\{ W(n)\right\} $. We now estimate
the effect of a small perturbation $\Delta t_{n},\ldots,\Delta t_{n-W}$
of the event times $t_{n},\ldots,t_{n-W}$ on the hypothetical event
time $t^{\prime}$. By Eq.~\eqref{eq:tprime}, the Jacobian of $t^{\prime}$,
$Dt^{\prime}$, with respect to former spike times, $t_{n},\ldots,t_{n-W}$,
is given as\begin{equation}
Dt^{\prime}\left(A\right)=\left(\frac{\partial t^{\prime}}{\partial t_{n}}\left(A\right),\ldots,\frac{\partial t^{\prime}}{\partial t_{n-W}}\left(A\right)\right)=-\left(\frac{\partial V_{i,n}}{\partial t}\left(A,t^{\prime}\right)\right)^{-1}\cdot DV_{i,n}\left(A,t^{\prime}\right).\end{equation}
The linearized estimation of the displacement $\Delta t^{\prime}$
of $t^{\prime}$ is then given by\begin{equation}
\Delta t^{\prime}\doteq Dt^{\prime}\left(A\right)\cdot\left(\begin{array}{c}
\Delta t_{n}\\
\vdots\\
\Delta t_{n-W}\end{array}\right)=\left(\frac{\partial V_{i,n}}{\partial t}\left(A,t^{\prime}\right)\right)^{-1}\cdot\sum_{k=n-W}^{n}\left(-\frac{\partial V_{i,n}}{\partial t_{k}}\left(A,t^{\prime}\right)\right)\cdot\Delta t_{k}.\label{eq:Deltatprime}\end{equation}
The special structure of $V_{i,n}(t)$ (cf.~Eq.~\eqref{eq:Vin}),
more precisely the fact that $V_{i,n}(t)$ depends on $t$ via $t-t_{k}$
for $k\in\left\{ n-W,\ldots,n\right\} $, yields the identity\begin{equation}
\sum_{k=n-W}^{n}-\frac{\partial V_{i,n}}{\partial t_{k}}\left(A,t^{\prime}\right)=\frac{\partial V_{i,n}}{\partial t}\left(A,t^{\prime}\right).\label{eq:identityofderivatives}\end{equation}
Under the condition,\begin{equation}
\frac{\partial V_{i,n}}{\partial t_{k}}\left(A,t^{\prime}\right)\leq0\quad\txt{for\, all\,}k=n-W,\ldots,n,\label{eq:condition}\end{equation}
we can combine Eq. \eqref{eq:Deltatprime} and Eq. \eqref{eq:identityofderivatives}
and find bounds for the the displacement \begin{equation}
\min_{k=\left\{ n-W,\ldots n\right\} }\Delta t_{k}\leq\Delta t^{\prime}\leq\max_{k=\left\{ n-W,\ldots n\right\} }\Delta t_{k}.\label{eq:tprimebounds}\end{equation}
Condition \eqref{eq:condition} implies that if neuron $i$ sends
or receives a spike earlier, also the threshold is crossed earlier.
This always holds for $\delta$-couplings, for interactions with temporal
extend it restricts the class of patterns as we show below. Eq.~\eqref{eq:tprimebounds}
is an analog to Eq.~\eqref{eq:DeltaBound}, sufficiently small perturbations
stay bounded by the initial ones for finite times. This directly implies
Lyapunov stability for periodic orbits. For general irregular dynamics
and to prove asymptotic stability, the propagation of pulses through
the network has to be studied as for the nonlinear stability analysis
in the main part.

We now want to specify a class of periodic patterns which are stable
in a network with temporally extended synaptic currents. The influence
of various events on $V_{i,n}(A,t^{\prime})$ is as follows: For an
influential spike receiving $t_{k}$, Eq.~\eqref{eq:Vin} yields\begin{equation}
\frac{\partial V_{i,n}}{\partial t_{k}}\left(A,t^{\prime}\right)=\frac{\beta}{\beta-\gamma}\eps^{\ast}\left(\gamma e^{-\gamma(t^{\prime}-t_{k})}-\beta e^{-\beta(t^{\prime}-t_{k})}\right),\label{eq:CondA}\end{equation}
where $\eps^{\ast}<0$ is the coupling strength from the sending neuron.
For the last spike sending of neuron $i$, $t_{k}=t_{0}(n,i)$, \begin{equation}
\frac{\partial V_{i,n}}{\partial t_{0}(n,i)}\left(A,t^{\prime}\right)={-I}_{i}e^{-\gamma(t-t_{0}(n,i))}<0.\label{eq:CondB}\end{equation}
For any other event $t_{k}$ a displacement of $t_{k}$ has no influence
on $V_{i,n}(t)$, here \begin{equation}
\frac{\partial V_{i,n}}{\partial t_{k}}\left(A,t^{\prime}\right)=0.\label{eq:CondC}\end{equation}
Therefore condition \eqref{eq:condition} reduces to a condition on
the left and right hand side of Eq.~\eqref{eq:CondA} and can be
reformulated as\begin{equation}
t^{\prime}-t_{k}>\frac{1}{\beta-\gamma}\ln\left(\frac{\beta}{\gamma}\right):=T_{d},\end{equation}
where $t_{k}$ are the spikes arrival times at neuron $i$ since the
last reset $t_{0}(n,i)$. This means that the class of patterns where
each neuron $i$ does not cross the threshold for a time period $T_{d}$
after receiving a spike are stable. For $\beta\rightarrow\infty$
the system tends to the $\delta$ - pulse coupled system and indeed
$T_{d}$ vanishes, $\lim_{\beta\rightarrow\infty}T_{d}=0$, such that
any non-degenerated orbit is stable. However, for temporally extended interactions
unstable periodic orbits exists and also chaotic dynamics is possible
(cf. Fig.~\ref{fig:extended}(A)).

To illustrate our analytical findings, in Fig~\ref{fig:extended}(B,C),
we used a generalization of a recently introduced method \citep{Memmesheimer2006,Memmesheimer2006a}
to design two networks realizing predefined spike patterns in a network
of five neurons ($V_{\Theta,i}\equiv1.0$, $I_{i}\equiv2.4$, $\beta=8$, $\tau:=\tau_{ij}\equiv0.125$) with temporally extended couplings. Both
patterns are the same, but with different inter-spike-intervals. In
(B) all spikes are separated by ${\Delta T=T}_{d}+\tau$, which ensures
that a neuron never spikes within a time period $T_{d}$ after receiving
a spike; in (C) we choose the inter-spike-intervals smaller $\Delta T=\left(T_{d}+\tau\right)/2$.
The lower panels illustrates the stability properties: The spike times
of the different neurons are plotted relative to the spike time of
neuron $1$ in vertical direction. The horizontal direction is simulated
time, different colors indicate spike times of the five different
neurons. At certain points in time (blue arrows) the network dynamics
is perturbed. The dynamics in (B) is stable: After perturbations of
size $\approx0.2$ (maximum norm), the dynamics converge towards the
periodic orbit. (C) The dynamics is unstable: a perturbation of size
$\approx10^{-12}$ leads to a divergence from the unstable periodic
orbit.

\section{Discussion}

Irregular spiking activity that robustly arises in balanced state
models, constitutes a generic feature of cortical dynamics. Here,
for a class of models including e.g.\ the common leaky integrate-and-fire
neuron, we have shown that generic trajectories which give rise to
the irregular balanced state can be exponentially stable. In particular, they
are stable in purely inhibitory
strongly connected networks of neurons with delayed couplings and
with infinitesimal synaptic time course. We numerically illustrated
and refined our analytical results. For small sparse networks we showed
that the dynamics even converges to a periodic orbit. However, the
length of the irregular yet stable transient grows rapidly with
network size such that for larger networks, in particular for biological relevant sizes, transients dominate the dynamics on all relevant time scales. 

Furthermore, we found that the phenomenon of stable yet irregular
dynamics is robust against introducing some excitatory interactions
or against increasing the synaptic time scales from zero. If the synaptic
responses become too slow or excitatory interactions too many, we
revealed a smooth transition from stable irregular dynamics to chaotic,
equally irregular dynamics. We emphasize that we kept the network
rate during this transition (and thus keep the balance) and that the
mean field descriptions \citep{Vreeswijk1998,Brunel2000} of networks
in both regimes are identical when the parameters are suitably chosen.
Thus, highly irregular spiking dynamics occurs independent of the stability
properties of the network. 

Earlier studies on balanced neural activity considered a priori the limit of infinitely many neurons in sparse networks \citep{Vreeswijk1996, Amit1997, Vreeswijk1998, Brunel1999, Brunel2000}. In this mean field limit the collective dynamics is well understood. In particular in infinitely large networks of binary neurons with balanced excitatory and inhibitory interactions the dynamics are chaotic \citep{Vreeswijk1996,Vreeswijk1998}. Further studies of finite networks found stable dynamics in weakly diluted networks of inhibitory coupled neurons \citep{Zillmer2006}, as well as in globally coupled   networks with dominating inhibition \citep{Jin2002}. Recent analytical evidence confirmed the existence of stable dynamics in inhibitory coupled networks of integrate-and-fire neurons with a more complex structure \citep{Jahnke2008}. As the inter-event times that underly our analysis shrink inversely proportional to the network size (at a given individual neuron-spiking-rate), the methods applied here, however, are not applicable in a straightforward way in associated mean field models. Thus, one cannot make strict statements about stability in the limit of infinitely many neurons. Nevertheless, as shown above, generic transients and periodic trajectories in arbitrarily large inhibitory coupled networks are stable.

Taken together, the results show that the microscopic dynamics in
 purely inhibitory coupled networks differs substantially
from the dynamics of networks that explicitly include
excitatory couplings. Whereas the latter is chaotic, the former is
completely different and generically stable - despite both showing the
same irregularity features. In particular, chaotic as well as stable
dynamics are equally well capable of generating highly irregular spiking
activity. The smoothness of the transition to chaos, without essential
change of the irregularity (e.g.~of the large coefficients of variation)
further suggests that chaos is not the main dynamical origin of the
high irregularity. We thus suggest that a mechanism different from
chaos contributes to the irregularity of cortical firing patterns
in a substantial way. Moreover, the location of the transition from chaos to stability as well as the dynamical mechanism underlying this transition, remain unknown.

Nevertheless, chaos as well as stochastic network properties such
as unreliable synapses, may support the robust occurrence of irregular
activity in cortical networks and also modify its computational features.
It is thus an important future task to investigate which anatomical
and dynamical features of cortical networks are indeed of crucial
relevance for their spiking activity and their functions.
\begin{acknowledgments}
We thank Fred Wolf for helpful comments, David Hansel for pointing
out the article \citep{Zillmer2009} and the Federal Ministry of Education
and Research (BMBF) Germany for partial support under Grant No. 01GQ430.
RMM thanks the Sloan Swartz Foundation for partial support.
\end{acknowledgments}

\section*{Conflict of Interest Statement}
The authors declare that the research was
conducted in the absence of any commercial or financial relationships that could be construed as a
potential conflict of interest.

\section*{Appendix}

In the following we will derive the bounds \eqref{eq:BoundsofDmxK},
\eqref{eq:BoundsOfDmnK} and \eqref{eq:BoundsCAst} stated in section
\ref{sec:asymptotic-stability}. Therefore, we will track the propagation of the perturbation of
one specific neuron $l_{0}$ through the entire network.

\begin{figure}
\begin{centering}
\includegraphics[clip,width=85mm]{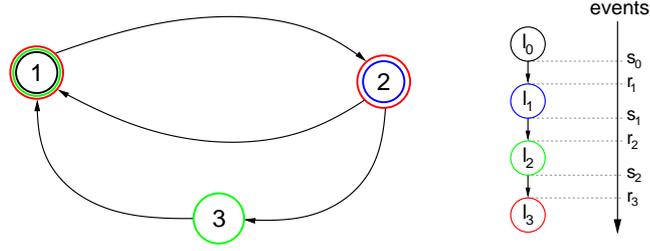}
\par\end{centering}

\caption{Tracking the propagation of a pulse. Here neuron $l_{0}=1$
(black) is fixed as the initial neuron. The sets of postsynaptic neurons
are: $\Post(1)=\left\{ 2\right\} $~(blue), $\Post^{2}(1)=\left\{ 1,3\right\} $~(green),
$\Post^{3}(1)=\left\{ 1,2\right\} $~(red). Following one specific
path through the network, we label the first spike event of neuron
$l_{i}\in\Post^{i}(l_{0})$ after receiving an input from $l_{i-1}\in\Post^{i-1}(l_{0})$
$s_{i}$. The event when the generated spike is received by $l_{i+1}$
is labeled $r_{i+1}.$ In a strongly connected network of size $N$
the union $\bigcup_{i=0}^{N-1}\left(\Post^{i}(l_{0})\right)=\left\{ 1,\ldots,N\right\} $
contains all the neurons, because any two of them are connected by
a directed path of maximal length $N-1$. \label{fig:tracking}}

\end{figure}

All neurons spike at least once in a sufficiently large but finite
time interval $T$. Moreover, after $\tau_{\txt{max}}=\max_{i,j}\left(\tau_{ij}\right)$
all spikes in transit have certainly arrived at the postsynaptic neurons.
We label the maximal number of events possible in the time interval
$\left[t,t+\max\left\{ T,\tau_{\txt{max}}\right\} \right]$ by $M$.
For purely inhibitory networks, $M<\infty$ due to the bounded neural
spike rate. We denote the set of postsynaptic neurons of $l_{0}$
by \begin{equation}
\Post^{i}(l_{0}):=\underbrace{\Post\circ\Post\circ\ldots\circ\Post(l_{0})}_{i\ \txt{times}},\label{eq:DefOfli}\end{equation}
thus a neuron $l_{i}\in\Post^{i}(l_{0})$ is connected to $l_{0}$
by a directed path of length $i$ (cf.~Fig.~\ref{fig:tracking}).
Further, we define $\Post^{0}(l_{0}):=\left\{ l_{0}\right\} $.

We estimate the bounds of the perturbation following one specific
path from a neuron $l_{0}$ to a neuron $l$. In a strongly connected
network, $l\in\Post^{j}(l_{0})$ for some $j\le N-1$, so there is
a directed path between $l_{0}$ and $l=l_{j}$ via neurons $l_{1},\ldots,l_{j-1}$
in the network. As the consideration holds for an arbitrary path,
the result is an universal bound of the perturbation. Initially, at
$n=0$, the neurons are perturbed by $\Delta_{i}^{(0)}$. The first
spiking of neuron $l_{0}$ after $n=0$ is labeled by $s_{0}\leq M$.
After a delay time $\tau_{l_{1}l_{0}}$ this spike is received by
the postsynaptic neuron $l_{1}\in\Post(l_{0})$, we call the event
$r_{1}\leq2M$. After at most $M$ further events, at $s_{1}\leq3M$,
the neuron $l_{1}$ emits a spike. In general, we recursively define
$s_{i}$ as the first spiking event of neuron $l_{i}\in\Post^{i}(l_{0})$
after $r_{i}$ and $r_{i}$ as the event when the spike generated
by $l_{i-1}\in\Post^{i-1}(l_{0})$ at $s_{i-1}$ is received (cf.~Fig.~\ref{fig:tracking}).
Due to the definition of $M$, the relations $s_{i}\leq(2i+1)M$ and
$r_{i}\leq2iM$ hold. 

First we prove by induction that the perturbation of the neuron $l_{i}$
before sending of a spike at $s_{i}$ is bounded from above by

\begin{equation}
\Delta_{l_{i}}^{(s_{i}-1)}\leq\left[\left(1-\cmx\right)^{i}\cdot\cmn^{s_{i}-1}\right]\Delta_{l_{0}}^{(0)}+\left[{1-\left(1-\cmx\right)}^{i}\cdot\cmn^{s_{i}-1}\right]\Dmx^{(0)}.\label{eq:IndStatement}\end{equation}

\begin{enumerate}
\item Initially neuron $l_{0}$ is perturbed by $\Delta_{l_{0}}^{(0)}$.
Before $l_{0}$ generates a spike it receives at most $s_{0}-1$ inputs.
According to Eq.~\eqref{eq:Averaging}, if the neuron $l_{0}$ indeed
receives an input the perturbation of neuron $l_{0}$ may increase.
To find an upper bound, we assume that at every event $0<n<s_{0}$
neuron $l_{0}$ receives an input with the maximal initial perturbation,
$\Dmx^{(0)}$, and a minimal averaging constant $\cmn,$ which moves
the average into the direction of the maximal possible perturbation.
Repeated application of \eqref{eq:Averaging} yields\begin{eqnarray}
\Delta_{l_{0}}^{(1)} & \leq & \cmn\Delta_{l_{0}}^{(0)}+\left(1-\cmn\right)\Dmx^{(0)},\nonumber \\
\Delta_{l_{0}}^{(2)} & \leq & \cmn\Delta_{l_{0}}^{(1)}+\left(1-\cmn\right)\Dmx^{(0)}\leq\cmn^{2}\Delta_{l_{0}}^{(0)}+\left(1-\cmn^{2}\right)\Dmx^{(0)}\nonumber \\
 &  & \ldots\nonumber \\
\Delta_{l_{0}}^{(s_{0}-1)} & \leq & \cmn^{s_{0}-1}\Delta_{l_{0}}^{(0)}+\left(1-\cmn^{s_{0}-1}\right)\Dmx^{(0)}.\label{eq:Bound_l_0}\end{eqnarray}
which is the inductive statement \eqref{eq:IndStatement} for $i=0.$
\item We assume that the statement \eqref{eq:IndStatement} holds for $\Delta_{l_{i}}^{(s_{i}-1)}$,
which is neuron $l_{i}$'s perturbation as inherited by the spike
sent at $s_{i}$ (cf.~Eq.~\eqref{eq:ShiftOfnewSpike}). After at
most $M$ events the spike is received by the postsynaptic neuron
$l_{i+1}$ at event $r_{i+1}$. In our worst- (or worse than worst-)
case estimation, we assume that neuron $l_{i+1}$ is maximally perturbed
before it receives the spike, $\Delta_{l_{i+1}}^{(r_{i+1}-1)}=\Dmx^{(0)},$
and that the interaction factor $c_{l_{i+1}}^{(r_{i+1})}$ is maximal,
$\cmx$, such that again the average is moved into the direction of
the maximal perturbation. Therefore the perturbation after the interaction
is bounded by \begin{eqnarray}
\Delta_{l_{i+1}}^{{(r}_{i+1})} & \leq & \cmx\cdot\Dmx^{(0)}+\left(1-\cmx\right)\Delta_{l_{i}}^{(s_{i}-1)}\nonumber \\
 & \leq & \left[\left(1-\cmx\right)^{i+1}\cdot\cmn^{s_{i}-1}\right]\cdot\Delta_{l_{0}}^{(0)}+\left[{1-\left(1-\cmx\right)}^{i+1}\cdot\cmn^{s_{i}-1}\right]\cdot\Dmx^{(0)}.\label{eq:Bound_l_i1_r_i}\end{eqnarray}
Before $s_{i+1}>r_{i+1}>s_{i}$, neuron $l_{i+1}$ receives at most
${(s}_{i+1}-1-r_{i+1})$ inputs. Analogously to Eq.~\eqref{eq:Bound_l_0},
we assume that with each event $l_{i+1}$ receives a spike which is
maximally perturbed (with $\Dmx^{(0)}$), and the averaging constant
is minimal, $\cmn$. This yields\begin{equation}
\Delta_{l_{i+1}}^{(s_{i+1}-1)}\leq\left[\left(1-\cmx\right)^{i+1}\cdot\cmn^{s_{i}-1+s_{i+1}-1-r_{i+1}}\right]\Delta_{l_{0}}^{(0)}+\left[{1-\left(1-\cmx\right)}^{i+1}\cdot\cmn^{s_{i}-1+s_{i+1}-1-r_{i+1}}\right]\Dmx^{(0)}.\label{eq:Bound_l_i1_a}\end{equation}
We replace $\cmn^{s_{i}-1+s_{i+1}-1-r_{i+1}}$ by $\cmn^{s_{i+1}-1}$
in Eq.~\eqref{eq:Bound_l_i1_a}, thereby increasing the right-hand
side, because $s_{i}-1-r_{i+1}<0$. This directly yields the induction
statement for $\Delta_{l_{i+1}}^{(s_{i+1}-1)}.$
\end{enumerate}
Based on Eq.~\eqref{eq:IndStatement} we now derive an upper bound
of the perturbation of all neurons after event $s_{N-1}$. After this
event every neuron has sent at least one spike which is influenced
by the initial perturbation of neuron $l_{0}$, because in a strongly
connected network the union $\bigcup\limits _{i=0}^{N-1}\Post^{i}(l_{0})$
contains all neurons of the network (cf.~Fig.~\ref{fig:tracking}).
After the $s_{i}$th event, neuron $l_{i}$ can still receive spikes.
Before the $s_{N-1}$th event, taken as reference, it receives in
the worst case scenario $(s_{N-1}-s_{i})$ inputs with maximal initial
perturbation $\Dmx^{(0)}$ and minimal averaging factor $\cmn$. Using
Eq.~\eqref{eq:IndStatement} we repeatedly apply Eq.~\eqref{eq:Averaging}
$(s_{N-1}-s_{i})$ times which leads to \begin{eqnarray}
\Delta_{l_{i}}^{(s_{N-1})} & \leq & \left[\left(1-\cmx\right)^{i}\cdot\cmn^{s_{i}-1+s_{N-1}-s_{i}}\right]\Delta_{l_{0}}^{(0)}+\left[{1-\left(1-\cmx\right)}^{i}\cdot\cmn^{s_{i}-1+s_{N-1}-s_{i}}\right]\Dmx^{(0)}.\label{eq:BoundafterSn1}\end{eqnarray}
The right-hand side increases with $i$, therefore the perturbation
of an arbitrary neuron $j\in\left\{ 1,\ldots,N\right\} $ after $s_{N-1}$
events is bounded from above by\begin{equation}
\Delta_{j}^{\left(s_{N-1}\right)}\leq\left[\left(1-\cmx\right)^{N-1}\cdot\cmn^{\left(s_{N-1}-1\right)}\right]\cdot\Delta_{l_{0}}^{(0)}+\left[{1-\left(1-\cmx\right)}^{N-1}\cdot\cmn^{\left(s_{N-1}-1\right)}\right]\cdot\Dmx^{(0)}.\label{eq:BoundOfAllNeurons}\end{equation}
At the $s_{N-1}$th event there can be $D^{\prime}$ spikes per neuron
in transit which are, in the worst-case scenario, assumed to have
the maximal perturbation. Due to their arrival after the $s_{N-1}$th
event, the perturbations of neurons can still increase. However, after
$s_{N-1}+M$ events all spikes generated before the $s_{N-1}$th event
have arrived at the corresponding postsynaptic neurons. Taking into
account the arrival of these spikes using Eqs.~(\ref{eq:Averaging},\ref{eq:BoundOfAllNeurons})
, we find an upper bound for the perturbation after $s_{N-1}+M$ events,\begin{equation}
\Delta_{j}^{\left(s_{N-1}+M\right)}\leq\left[\left(1-\cmx\right)^{N-1}\cdot\cmn^{\left(s_{N-1}-1+M\right)}\right]\cdot\Delta_{l_{0}}^{(0)}+\left[{1-\left(1-\cmx\right)}^{N-1}\cdot\cmn^{\left(s_{N-1}-1+M\right)}\right]\cdot\Dmx^{(0)}.\label{eq:BoundOfAllNeuronsAfterArrival}\end{equation}
Due to the fact that generated spikes inherit a perturbation present
in the phases at spike sending time, the bound \eqref{eq:BoundOfAllNeuronsAfterArrival}
holds also for the perturbation of spikes generated after the $s_{N-1}$th
and before the $(s_{N-1}+M)$th event, because the bound \eqref{eq:BoundOfAllNeuronsAfterArrival}
limits the maximal perturbation for all neurons between the $s_{N-1}$th
and the $(s_{N-1}+M)$th event.

 We conclude that the perturbations of the neurons and the spikes
in transit after $K:=2NM\ge s_{N-1}+M$ events are bounded by\begin{equation}
\Delta_{j}^{(K)}\leq\left[\left(1-\cmx\right)^{N-1}\cdot\cmn^{2NM-1}\right]\cdot\Delta_{l_{0}}^{(0)}+\left[{1-\left(1-\cmx\right)}^{N-1}\cdot\cmn^{2NM-1}\right]\cdot\Dmx^{(0)}.\label{eq:BoundsAfterKEvents}\end{equation}
 Therefore we find an upper bound for the maximal perturbation $\Dmx^{(K)}$
after $K$ events,

\begin{equation}
\Dmx^{(K)}\leq c^{\ast}\cdot\Delta_{l_{0}}^{(0)}+\left[1-c^{\ast}\right]\cdot\Dmx^{(0)}.\label{eq:BoundsofDmxKApp}\end{equation}
with \begin{equation}
{0<c}^{\ast}:=\left(1-\cmx\right)^{N-1}\cdot\cmn^{2NM-1}\leq\left(1-\cmx\right)\cdot\cmx\leq1/4,\label{eq:DefOfCstar}\end{equation}
\begin{equation}
3/4\leq\left(1-c^{\ast}\right)<1.\label{eq:1minusCStar}\end{equation}
Similarly, we find a lower bound for the minimal perturbation after
$K$ events\begin{equation}
\Dmn^{(K)}\geq c^{\ast}\cdot\Delta_{l_{0}}^{(0)}+\left[1-c^{\ast}\right]\cdot\Dmn^{(0)},\label{eq:BoundsOfDmnKApp}\end{equation}
by an estimation analogous to the one above, where only $\Dmx^{(0)}$
has to be replaced $\Dmn^{(0)}$ and the relation {}``$\leq$'' has
to be replaced by {}``$\geq$''. We note, that we did not have to
specify the perturbation $\Delta_{l_{0}}^{(0)}$ to derive this result.


\end{document}